\documentclass{aastex}
\usepackage{times}
\usepackage{emulateapj5}

\def\be{\begin{equation}}
\def\ee{\end{equation}}

\newcommand{\labfig}[1] {\label{fig:#1}}
\newcommand{\fig}[1] {\ref{fig:#1}}

\newcommand{\mb}[1] {\mbox{\boldmath $#1$}}

\slugcomment{The Astrophysical Journal Supplement}

\lefthead{Barnes {\it et al.}}
\righthead{ApJS, December 2002}

\begin{document}

\title{The {\sl MAP} Satellite Feed Horns}
 
\author{Chris Barnes\altaffilmark{1}, Michele Limon\altaffilmark{1,2}, 
Lyman Page\altaffilmark{1}, 
Charles Bennett\altaffilmark{2}, Stuart Bradley\altaffilmark{1},\\ 
Mark Halpern\altaffilmark{3}, Gary Hinshaw\altaffilmark{2}, 
Norm Jarosik\altaffilmark{1}, William Jones\altaffilmark{1}, 
Al Kogut\altaffilmark{2}, \\
Stephan Meyer\altaffilmark{4}, Olexei Motrunich\altaffilmark{1}, \\
Greg Tucker\altaffilmark{6}, David Wilkinson\altaffilmark{1}, 
\& E. J. Wollack\altaffilmark{2}
}

\altaffiltext{1}{Dept. of Physics, Princeton University, Princeton, NJ 08544}
\altaffiltext{2}{Code 685, Goddard Space Flight Center, 
                 Greenbelt, MD 20771}
\altaffiltext{3}{Dept. of Physics, Univ. Brit. Col., Vancouver, B.C., 
                 Canada V6T 1Z4}
\altaffiltext{4}{Astronomy and Physics, University of Chicago, 
                 5640 South Ellis Street, LASP 209, Chicago, IL 60637}
\altaffiltext{5}{Dept of Astrophysical Sciences, Princeton University,
                 Princeton, NJ 08544}
\altaffiltext{6}{Dept. of Physics, Brown University, Providence, RI 02912}
\email{page@princeton.edu}


\begin{abstract}
We present the design, manufacturing methods, and characterization of
20 microwave feed horns currently in use on the Microwave Anisotropy
Probe (MAP) satellite.  The nature of the cosmic microwave background
(CMB) anisotropy requires a detailed understanding of the properties of
every optical component of a microwave telescope.
In particular, the properties of the feeds must be known so that the
forward gain and sidelobe response of the telescope can be modeled
and so that potential systematic effects may be computed.
{\sl MAP} requires low emissivity, azimuthally
symmetric, low-sidelobe feeds in five microwave bands (K, Ka, Q, V,
and W) that fit within a constrained geometry.  The beam pattern
of each feed is modeled and compared with measurements; the agreement
is generally excellent to the -60 dB level (80 degrees from the beam
peak).  This agreement verifies the beam-predicting software and the
manufacturing process. The feeds also affect the properties and modeling
of the microwave receivers. To this end, we show that the 
reflection from the feeds is less than -25~dB over most of each band
and that their emissivity is acceptable.
The feeds meet their multiple requirements.
\end{abstract}

\section{Introduction}
\label{section:intro}
The goal of the Microwave Anisotropy Probe ({\sl MAP}) satellite is to 
produce a high-fidelity
polarization-sensitive map of the microwave sky in five frequency bands
between 20 and 100 GHz \cite{Bennett03}. The primary science goal is to 
characterize the 
anisotropy in the cosmic microwave background (CMB). Maps of the sky are
produced from a set of differential measurements.  
Two mirror symmetric arrays of corrugated feeds
\cite{clarricoats,thomas} couple radiation from MAP's two telescopes to
the inputs of the differential receivers as shown in Figure 
1. Corrugated feeds were chosen because of their low emissivity,
symmetric beam pattern, low sidelobes, and because their pattern can
be accurately computed.  To the base of each feed is attached an
ortho-mode transducer (OMT)\footnote{Our OMT is a waveguide 
device with one dual-mode circular input and two rectangular output
ports. Ideally, the OMT splits the two input
orthogonal linear polarizations into
separate components and emits them into two rectangular waveguides. In
practice, there is always some reflection and mixing of modes.} 
one output of which is the input of one side of a 
differential radiometer. The feed's wide 
end opens into empty space, accepting radiation from the secondary 
mirror.

\begin{figure*}[tb]
\epsscale{1.3}
\plottwo{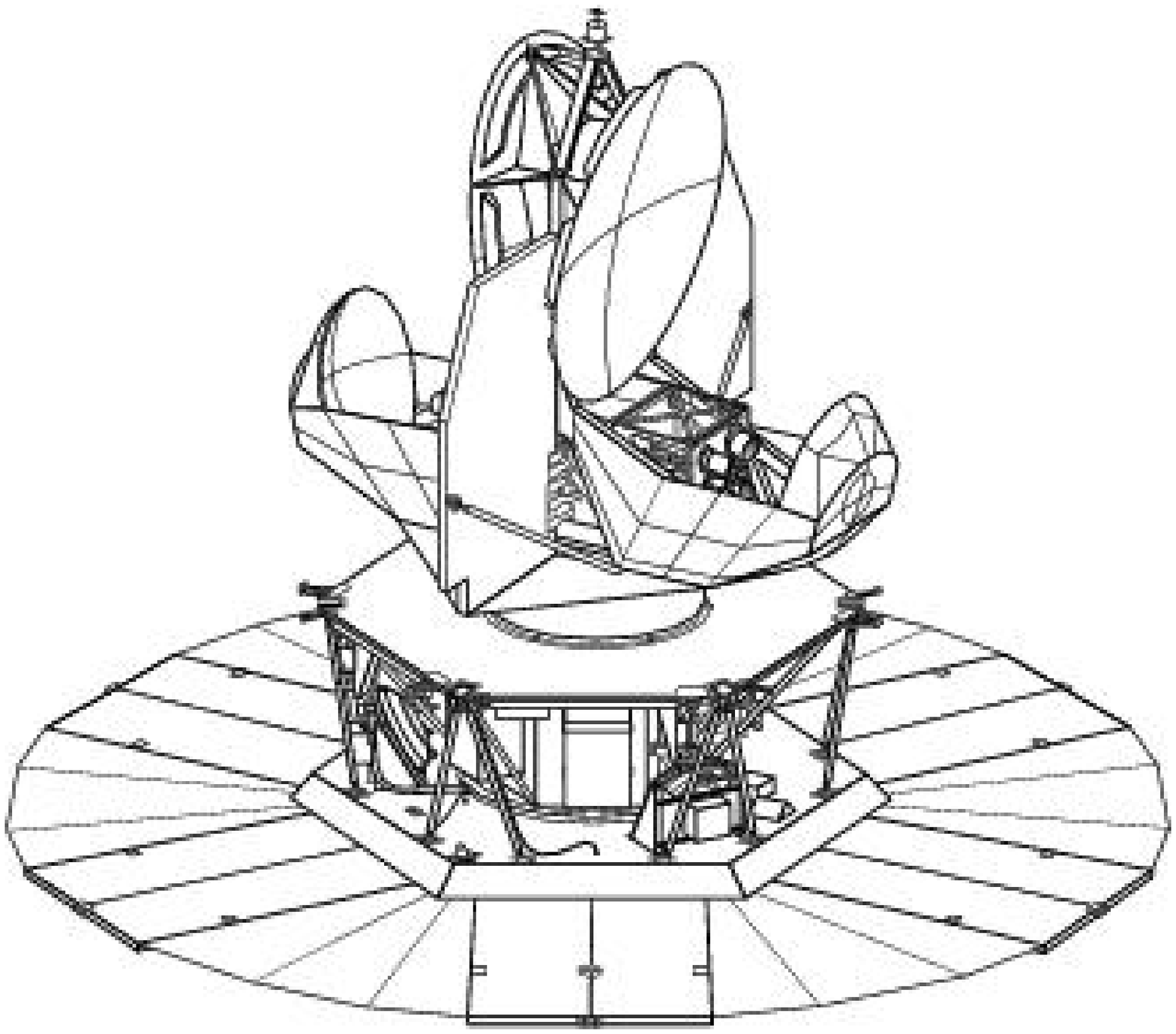}{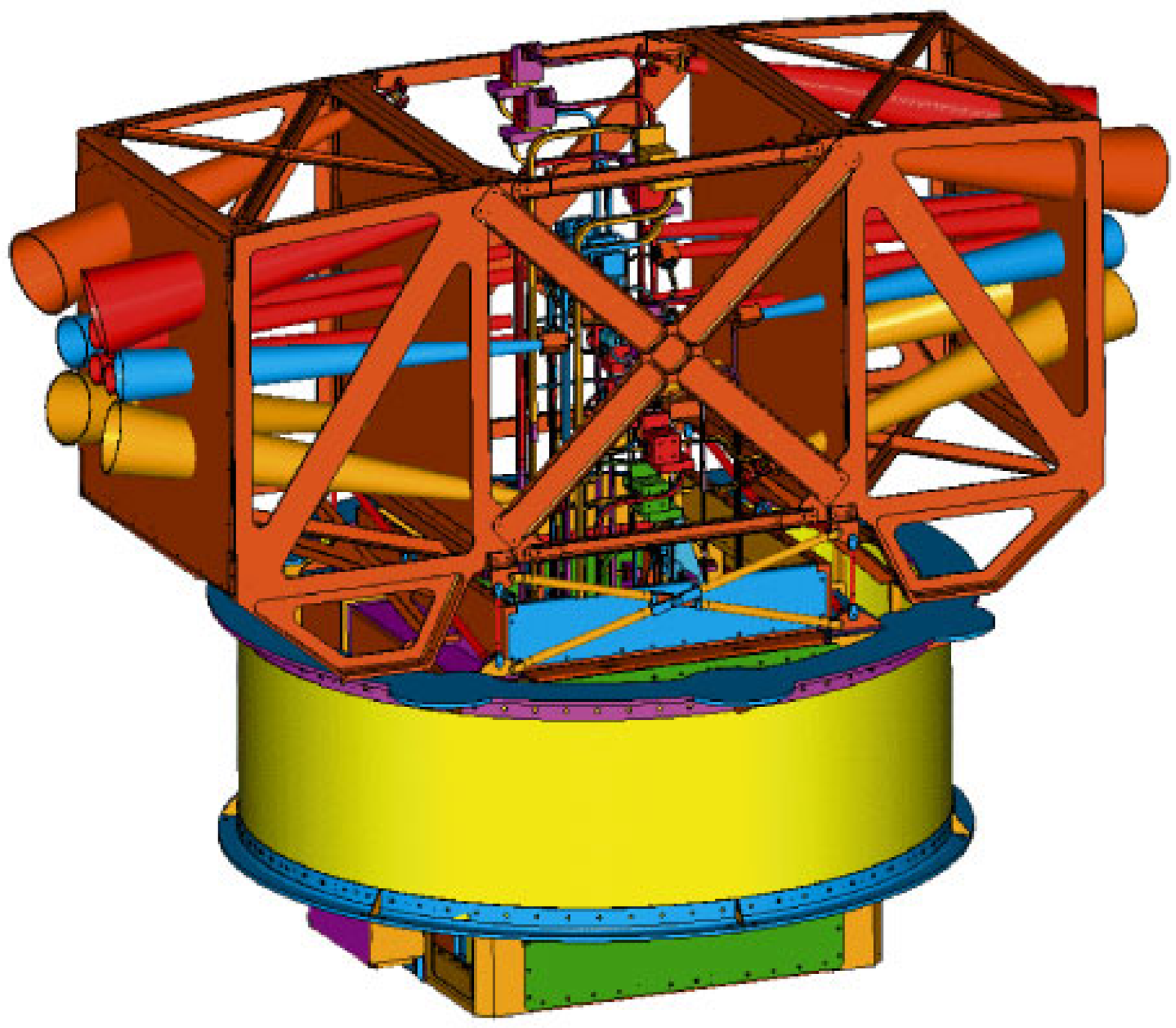}
\caption{\small{\it Left}: A line drawing of the {\sl MAP} satellite. Two
back-to-back shaped Gregorian telescopes focus incident radiation onto
the feeds which are located directly below the primary mirrors. The
large circular disk at the bottom shields the telescopes from sunlight.
The diameter of the disk is 5.1 m and the satellite is 3.6 m high. 
{\it Right}: The ``focal plane assembly'' (FPA) that houses the feeds.
There are ten feeds on each side. The central area of the FPA houses the
microwave receivers.}
\labfig{scandfeeds}
\end{figure*}   

The accuracy with which {\sl MAP} aims to measure the  
CMB anisotropy signal is $\sim 1 \mu K$, much smaller than
the brightest objects in the microwave sky. Thus, it is crucial to understand 
all contributions to the measured flux.
The {\sl MAP} feeds were designed to have a beam size of 
$\approx 9\deg$ full width
at half maximum (FWHM), illuminating the secondary mirrors roughly
equally in all bands. Beam spill past the secondary 
leads to an unfocused lobe offset from the main beam.
If the secondary subtended only a cone of half angle 
$20\deg$ as viewed from a feed, the Galactic center in Q-band would contribute a
$> 130\ \mu$K signal directly through a feed as it was illuminated from
over the rim of the secondary. The need
for properly shielding and directing the feed response is clear. 
From the center of the focal plane, the perimeter of the 
secondary is $40\deg$ from the primary optical axis.
The edge-taper on the secondary (ratio of the 
intensity at the edge to the
peak value in the middle) is $<-50~$dB for all bands. Just as important
as the shielding is the confidence that the far off-axis response of the feeds
is understood so that firm limits may be placed on contamination. 
{\sl MAP}'s primary beams are between $0\fdg2$ and 
$1\deg$ in FWHM though they are not Gaussian. The interplay between the feed 
response and that
of the full optical system is discussed in a companion paper \cite{Page03}.

\newpage
\section{Principles and Design of {\sl MAP's} Corrugated Feeds}
\label{section:principles}

The purpose of the {\sl MAP} feeds is to smoothly transform a single circular
TE${}_{11}$ mode in the OMT into a hybrid HE${}_{11}$ mode in the feed's
skyward aperture. The resulting beam is, as nearly as
possible, cylindrically symmetric, linearly polarized and Gaussian in
profile.
 
We may think of the feed in four separate sections.  First, the input
aperture (``input'') couples the feed to its OMT.  This section is
characterized by an input diameter and a short section of cylindrical
guide.  Second, the throat transforms the TE${}_{11}$ mode in the
circular guide into the hybrid mode HE${}_{11}$.  This section is
characterized by $\sim 10$ grooves ranging from $\lambda/2$ to
$\lambda/4$ deep.  The third is the corrugated waveguide that
propagates the hybrid mode. This section is characterized by grooves
of depth $\lambda/4$, and may be conical or profiled. (The radius of a
profiled horn grows nonlinearly along its length.)  The fourth is the output
which launches the hybrid mode.  The output is characterized by a diameter
and profile that can be shaped \cite{RRS} to fine tune the beam.

The input diameter is chosen to impedance match the standard waveguide
sizes and circular waveguide over the band. In particular, we match
the cutoff wavelength of the rectangular waveguide TE$_{10}$ mode to
the cutoff wavelength of the circular guide TE$_{11}$ mode and set
$d_{input}=2a_0(1.841/\pi)$, where $a_0$ is the broad-wall dimension of the
waveguide. 
Consequently, the circular waveguide impedance for TE$_{11}$
matches the wave impedance for TE$_{10}$ in rectangular waveguide: 
Z$_{TE_{10}}$ = $\eta k/k_g=~$ Z$^{circ}_{TE_{11}}$.
Here $k$ is the free-space wavenumber, $k_g$ is the
guide wavenumber, and $\eta $ is the wave impedance for a plane wave. 
This practice was followed successfully in 
Wollack {\it et al.} (1997).

The throat performs two separate functions: it
matches the impedance of the smooth circular input waveguide to the
grooved section of the horn, and ideally it transforms a TE$_{11}$ mode
into a HE$_{11}$ mode.  The feed throat design follows the 
work of Zhang (1993). A sample horn throat is shown in Figure \fig{mode_form}.

\begin{figure*}[tb]
\epsscale{0.9}
\plotone{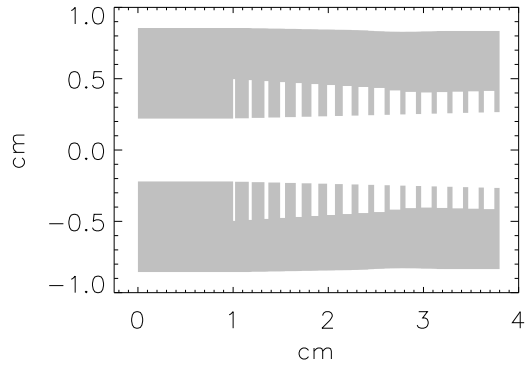}
\caption{\small The mode-forming section of a V-band horn with center frequency
60 GHz. The rightmost six grooves are $\lambda/4$ deep.  The OMT bolts
onto the left side, and supplies a right-traveling TE$_{11}$ mode.
The 1 cm long section of cylindrical waveguide ensures that the
TE$_{11}$ is well established.  Full groove patterns for all bands are
available on request.}
\labfig{mode_form}
\end{figure*}

Once the HE${}_{11}$ mode is formed, the corrugated horn expands
adiabatically outward, stretching the mode to fill a larger cylinder,
while maintaining its polarization, symmetry, and field pattern. The
goal is to flare the horn to shape the HE$_{11}$ mode into a
nearly Gaussian shape with a small cross-polar component.
As the feed flares, the axial wavenumber $k_g \to k_0$, the free space
value,  inside the $\lambda/4$
corrugated waveguide, so propagating fields experience only a small
discontinuity at the end of the feed. 
Accordingly, the propagating
mode does not reflect from the skyward aperture, nor does it 
significantly diffract
around the feed edges. (The edge currents, which are proportional to
the edge fields, are close to zero.) The mode detaches as
though it were already traveling through free space.  

The output diameter of the feed was set
following Touvinen (1992) to produce a 
$\approx 9\deg$ full width at half maximum ($\theta_{fwhm}$) beam. The aperture 
field for the HE$_{11}$ mode is given by
$
E_x(\rho)=0,\quad E_y(\rho) = A {J_0}^\top(k_c\rho) 
e^{-ik\rho^2/2R_{\rm sl}}\ ,
$
where $ {J_0}^\top(x) = J_0(x)$ for $x \le \xi$ and 0 otherwise. Here
$\xi \approx 2.405$ is the first zero of $J_0$, and $k_c = \xi/a$
where $a$ is the aperture radius, distinct from $R_{\rm sl}$, the
effective slant length of the cone. 

To first order in the far field, the emitted beam is Gaussian: \\
$E(\theta,z) \approx A \cos\theta e^{-\sin^2\theta (k w_0/2)^2}
e^{- ikz}\ , $ where $z$ is the direction along the axis of the feed,
$\theta$ is the angle away from the $z$ axis, and
$w_0 = \zeta a/\sqrt{1+\zeta^4(k a^2/2R_{\rm sl})^2}$ is a 
virtual beam waist size. The constant, $\zeta \approx
0.6436$, is a dimensionless and numerically determined by
maximizing power in the fundamental Gaussian mode.
For a narrow beam, the full width at half maximum is thus
$\theta_{fwhm} \approx 2 {\sqrt{2\ln 2} \over k w_0}.$
These considerations led to the specification of the aperture, though in
the final design the aperture diameter was optimized.

The final horns are shaped as in Figure \fig{horns}.  
Due to geometric constraints from the fairing diameter
and receiver assembly, all the feeds are nearly the same length.  This
required lengthening the Q, V and W band feeds beyond their natural
conical length \cite{thomas} and cosine-profiling the K-band feed
\cite{mahoud} to shorten it from its natural conic length.  
In addition to satisfying the length constraints, the 
Q, V, and W feeds are flared \cite{RRS} to 
enhance the Gaussanity of the beam. 

Some microwave
feeds have choke grooves around the rim of the aperture. These were
not necessary for {\sl MAP} because the gain of the feeds is high 
($\approx26~$dB); the
coupling of two side-by-side feed horns is predicted to be
$<-100$~dB.  With the full optical system assembled, the measured
coupling between pairs of the four feeds in W band is $-75$~dB to
$-80$~dB. This cross talk is due to reflections from the microwave
shielding around the secondary, not to direct feed-to-feed coupling,
and is still too small to generate any observable correlations between
radiometers.

\begin{figure*}[tb]
\epsscale{1.3}
\plotone{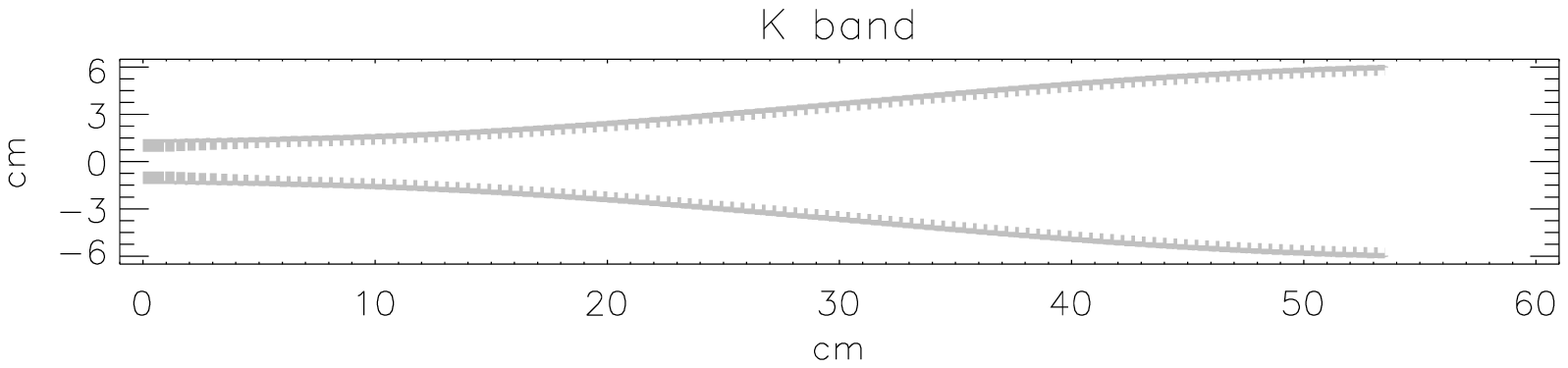}\\
\plotone{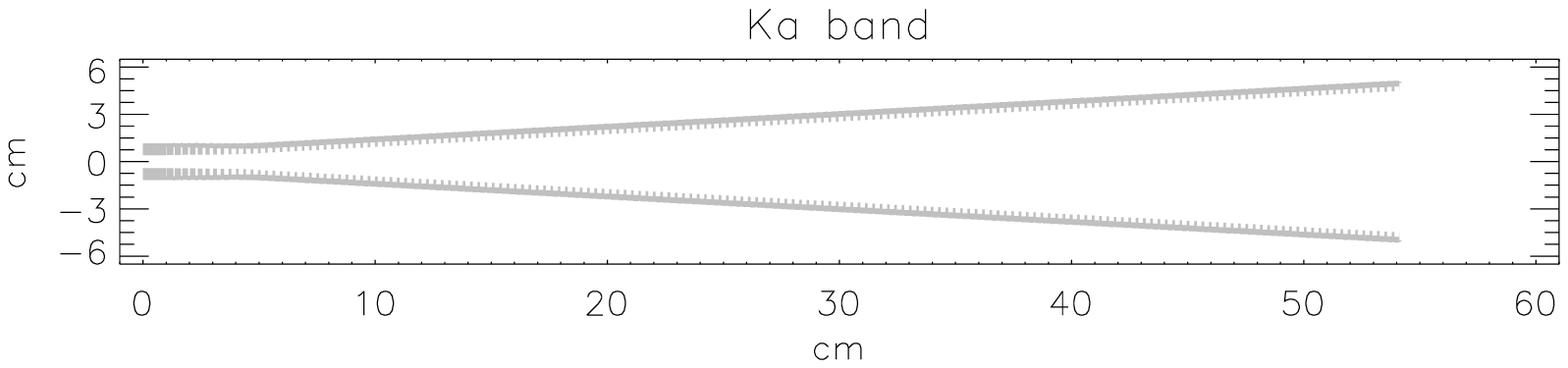}\\
\plotone{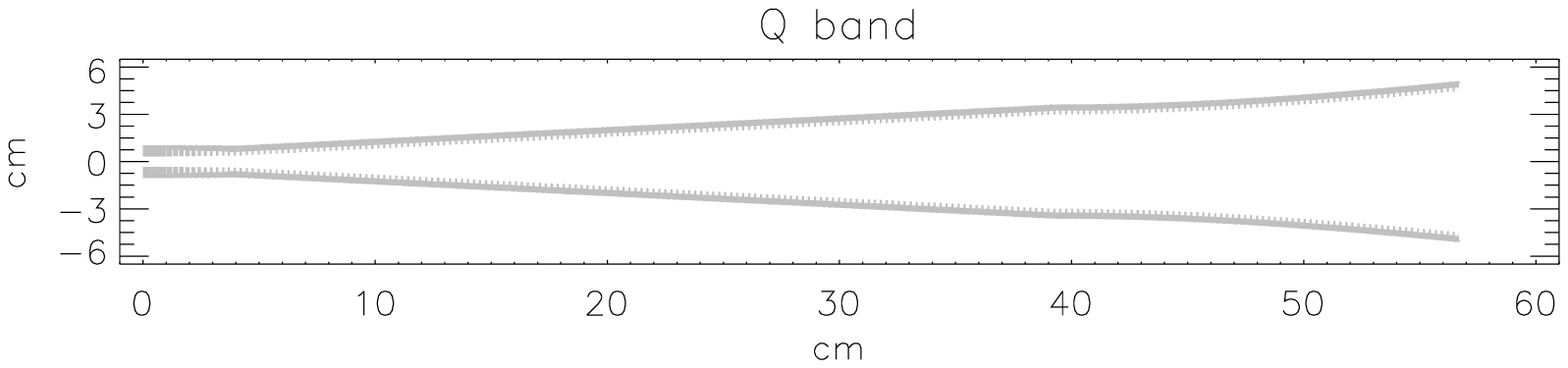}\\
\plotone{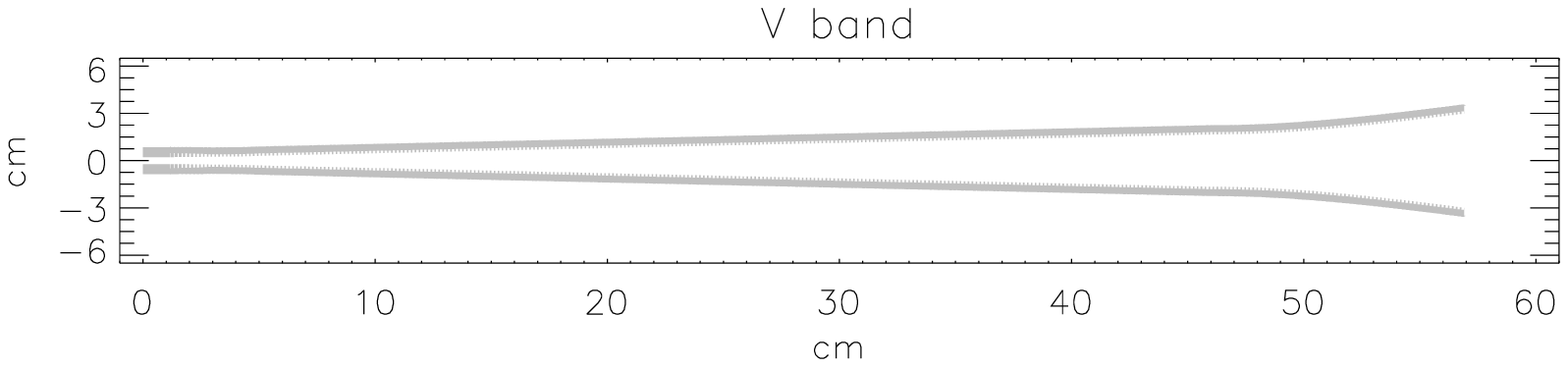}\\
\plotone{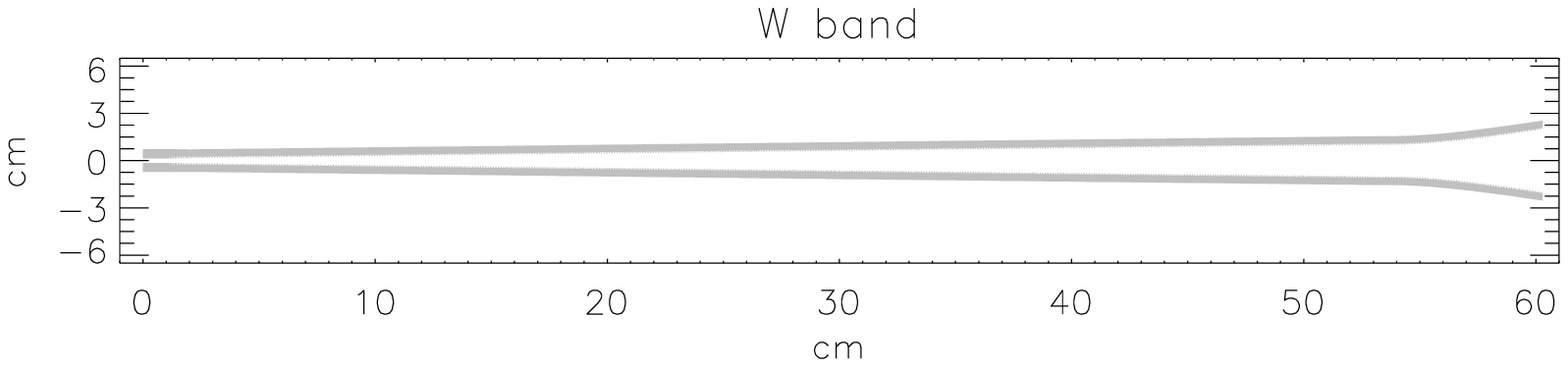}\\
\caption{\small Profiles of the {\sl MAP} feed horns.
The corrugations are clearly visible in the K-band feed and less visible
in the others due to printing resolution. The shapes and precise 
dimensions are available upon request.}
\labfig{horns}
\end{figure*}

\section{Fabrication of the Feeds}

\label{section:design}

Table
\ref{table:hornspecs} shows the specifications
for each band.  The K and Ka band horns are turned from single blocks of
7075 aluminum. 
The higher frequency {\sl MAP} feed horns: Q, V, and W
bands, are long and narrow and are therefore fabricated in 3, 4, and 5 
short sections respectively which bolt together. 
Each joint is placed on a groove boundary, so the two
sections mate to complete the groove; a cylindrical lock-and-key
pattern forces these flanges to self-align, as in Figure
\fig{joint}. The section closest to the OMT is electro-formed in copper, 
and then gold plated.  The remaining, wider sections are machined from
7075 aluminum because of its machineability.  Each machined 
section went through a regimen of
repeated inspection, 380~K to 77~K thermal cycles, ultrasound, air jet
and methanol cleaning to remove metal chips.  After cleaning the
measured interior surface roughness is $0.7\ \mu$m RMS.

\begin{table*}[t]
\caption{Dimensions of the {\sl MAP} feeds}
\small{
\vbox{
\tabskip 1em plus 2em minus .5em
\halign to \hsize {#\hfil &\hfil#\hfil &\hfil#\hfil &\hfil#\hfil  
                          &\hfil#\hfil &\hfil#\hfil &\hfil#\hfil 
                          &\hfil#\hfil &\hfil#\hfil &\hfil#\hfil 
                          &\hfil#\hfil &\hfil#\hfil \cr
\noalign{\smallskip\hrule\smallskip\hrule\smallskip}
Band & WR& $\nu$ & $\nu_{gv}$  & Length & Apt. & Mass &
Throat & $N_{gv}$ & $N_s$ & $\theta_{fwhm}$ & Gain \cr
&  & GHz & GHz & cm & cm & gm & cm & & &  deg & dBi \cr 
K  & 42 & 19.5--25 & 22.0 & 53.64 & 10.9374& 1010 & 1.2496 & 116 & 1 &
 10.1--7.7 & 24.9--28.1 \cr
Ka & 28 & 28--37   & 25.9 & 54.21 & 8.9916 & 650 & 0.8336 & 169 & 1 &
  8.9-6.7 &26.1--29.1\cr
Q  & 22 & 35--46   & 32.5 & 56.76 & 8.9878 & 615 & 0.6680 & 217 & 3 &
  7.8-6.0&27.3--30.8\cr
V  & 15 & 53--69   & 49.1 & 56.96 & 5.9890 & 325 & 0.4408 & 329 & 4 &
  8.8-7.4 &26.0--28.8\cr
W  & 10 & 82--106  & 90.1 & 60.33 & 3.9916 & 214 & 0.2972 & 533 & 5 &
  9.7-8.3 & 25.0--27.2\cr
\noalign{\smallskip\hrule}
}}}
\small{Diameters are 
given for the feed length, aperture and throat. $N_s$ is 
the number of 
separate machined pieces from which the horn is assembled. Standard
waveguide bands were chosen to minimize fabrication and testing
costs for the microwave receivers. $\nu_{gv}$ is the frequency for 
which the horn groove depth is $\lambda/4$ (sometimes called the 
hybrid frequency). The antenna beam width range
and gains run from the
lowest to highest frequencies in the band. (The number shown is the
average of E and H-plane FWHM.)}
\label{table:hornspecs}
\end{table*}

\begin{figure*}[tb]
\plotone{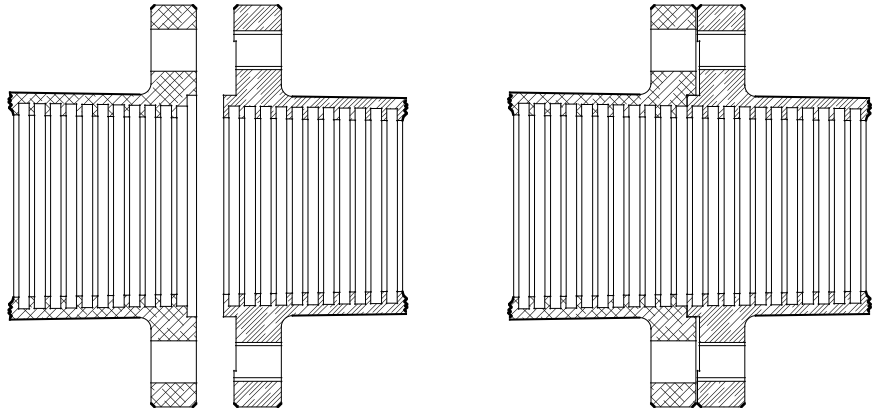}
\caption{\small A joint between two sections of a W-band feed. The
machining tolerance is 40~$\mu$m on the grooves and 5~$\mu$m on the 
joints. The flange design axially constrains the mating of the 
two section. Once the press fit joint is properly seated and the screws
uniformly torqued, the joint is robust.}
\labfig{joint}
\end{figure*}

The final stage of fabrication consists of joining the feed sections.
Especially at high frequencies, each joint must be mated carefully; a
gap or misalignment can cause reflections or launch unwanted modes.
At W-band, the feed joints are so critical and sensitive to proper
mating that the feeds were assembled joint by joint, with the
feed attached to a network analyzer. 
\footnote{ 
Because of this sensitivity, a W-band feed was mapped, vibrated at 
space qualification levels, and mapped again.  No deterioration was seen.  
Similarly, in the integration and testing phase, the balance and 
noise properties of the microwave radiometer were measured with cold
feeds.  The assembly was then warmed up, vibrated to space
qualification levels, cooled down, and re-measured. No change in
performance was seen.} 
Spikes in the reflection
spectrum often showed otherwise invisible flaws in the
section-to-section mate. This procedure entwined the processes of
building the horns and of measuring their properties.  

\smallskip

\section{Detailed predictions of antenna patterns, reflections and loss}
\label{section:predictions}

Once the horn design is known in detail, it is possible to calculate
its beam pattern precisely. {\sl MAP} uses a commercial program called
CCORHRN\cite{yrs}. The
algorithm solves Maxwell's equations exactly within the
feed \cite{james}, so its
predictions are correct up to manufacturing defects.
The result is a complete solution for the radiation field inside the feed. 
The full pattern at any point in space can then be computed from a
spherical wave expansion of the field in the skyward aperture. As the 
secondary mirrors are not in
the far field of the feeds ($> 8 a^2/\lambda$) such an expansion 
is required for accurate predictions of the full telescope pattern.

The same model may be used to compute the loss in the feed
\cite{clarricoats}. In the W-band feeds, most of the loss comes
from the tenth through fiftieth grooves in the horn, counting from the
OMT aperture. At 293 K, for a gold surface 
($\sigma=4.4\times10^7~$mhos/m ), the calculated emission temperature is 5.1~K
corresponding to an emissivity of $\epsilon={\rm
T_{emis}/T_{phys}}=0.017$. 
If one approximates the throat as a 1~cm
length of round guide followed by corrugated waveguide with grooves of
depth $\lambda/4$, the computed emissivity \cite{clarricoats} is
0.008, a factor of two less than for the full calculation.

\section{Measurements and Modeling of the Beam Patterns.}
\label{section:beams}

The antenna patterns of all {\sl MAP}'s feeds
were measured before installation into the satellite. While
a manufacturing flaw will typically alter both the horn's reflection
spectrum and its beam pattern, it is quite possible for a scratch,
metal chip, or miscut groove to alter one but not the other\footnote{A
defect which altered {\it neither\/} the feed's reflections nor its
beam would be transparent to us, and it would have no effect on the
telescope's performance.}.

Princeton's mathematics department is housed in Fine hall tower, a
thin rectangular building $\sim$ 50 m tall.  From the roof of the
physics department, $\sim$ 70 m away, the tower is silhouetted alone
against empty sky, which makes it an ideal place for a microwave
source.  We mount sources, one for each band ranging from 140 to 800
mW in strength, on top of 
the tower and direct their radiation,
modulated at 1 kHz, at the roof of the physics building with 
standard-gain horns.  As a result, a
$\sim$ 5m wide region of the physics department roof is illuminated
with uniform, monochromatic, unidirectional microwave radiation.

Both the E- and H-planes\footnote{For linearly
polarized radiation, the E (H) plane of a feed antenna is the plane of
the beam axis and electric (magnetic) field direction.  An E-plane 
beam measurement is an angular scan of the antenna response as the 
horn swings through orientations $\hat{\mb{n}} = \hat{\mb{s}} \cos\theta +
\hat{\mb{E}} \sin\theta$, where $\hat{\mb{s}}$ points toward the source, and 
$\hat{\mb{E}}$ is the electric field direction and $\theta$ is the
horn's angle from the source direction. In other words, the scan
direction is along the direction of polarization of the incident
radiation. The H-plane corresponds to the perpendicular scan direction.}
are measured with both vertical and horizontal scans. 
For the E-plane, first the E-plane co-polar response is 
determined, for example with a vertical scan. Next the 
source polarization is rotated 90 degrees and the opposite port on the 
OMT is selected and the measurement repeated, yielding another E-plane 
beam pattern but in a horizontal scan. The H-plane is measured similarly.
This double-scanning technique served to isolate unwanted geometrical
features on the range. For instance, any reflection that interferes with the 
H-plane beam pattern when one scans {\it sideways\/} would be unlikely
to contribute when the scan is {\it vertical\/} (and even less
likely to distort the antenna pattern in the same way).

\begin{figure*}[tb]
\epsscale{0.68}
\plotone{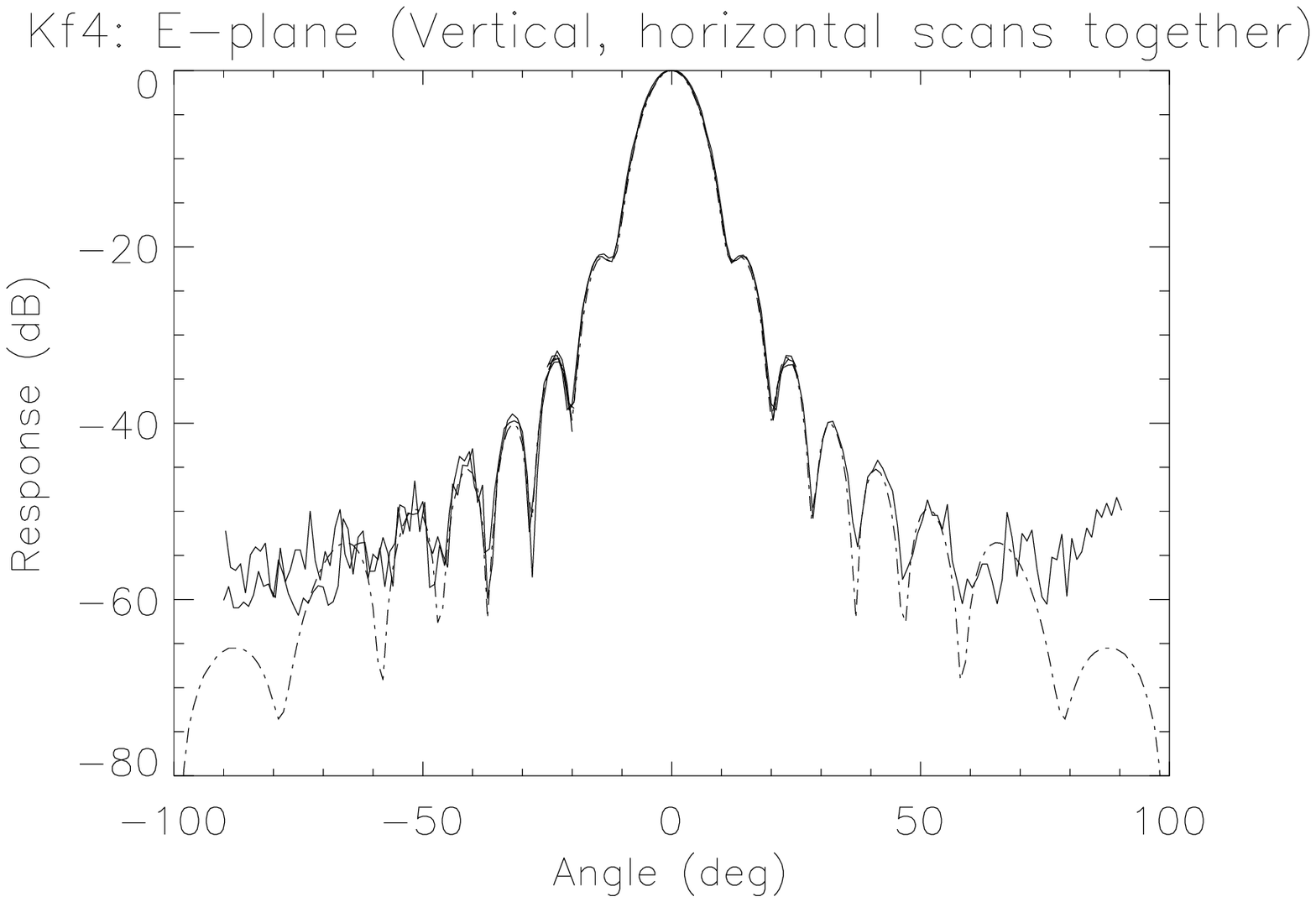} 
\plotone{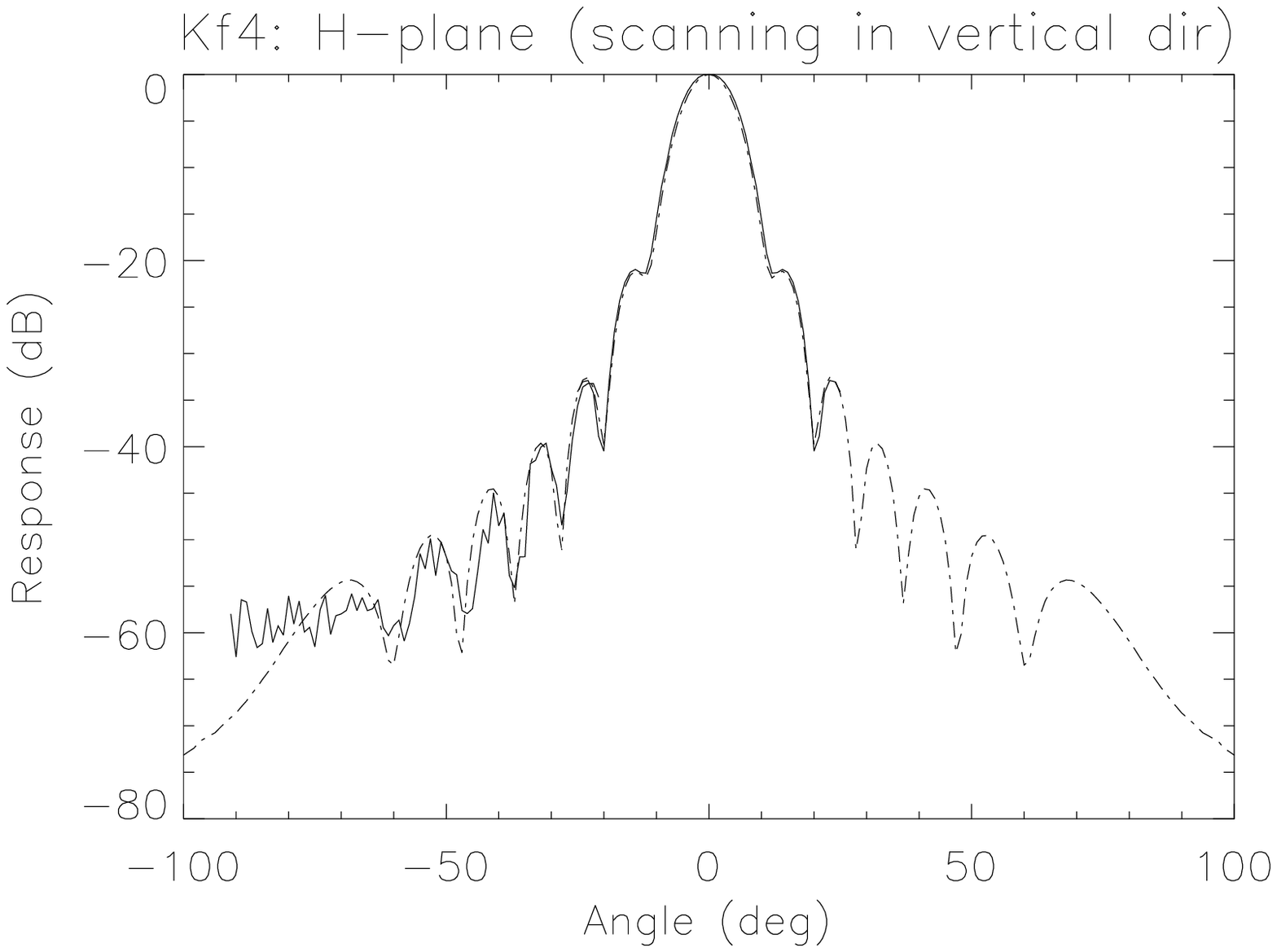} \\
\plotone{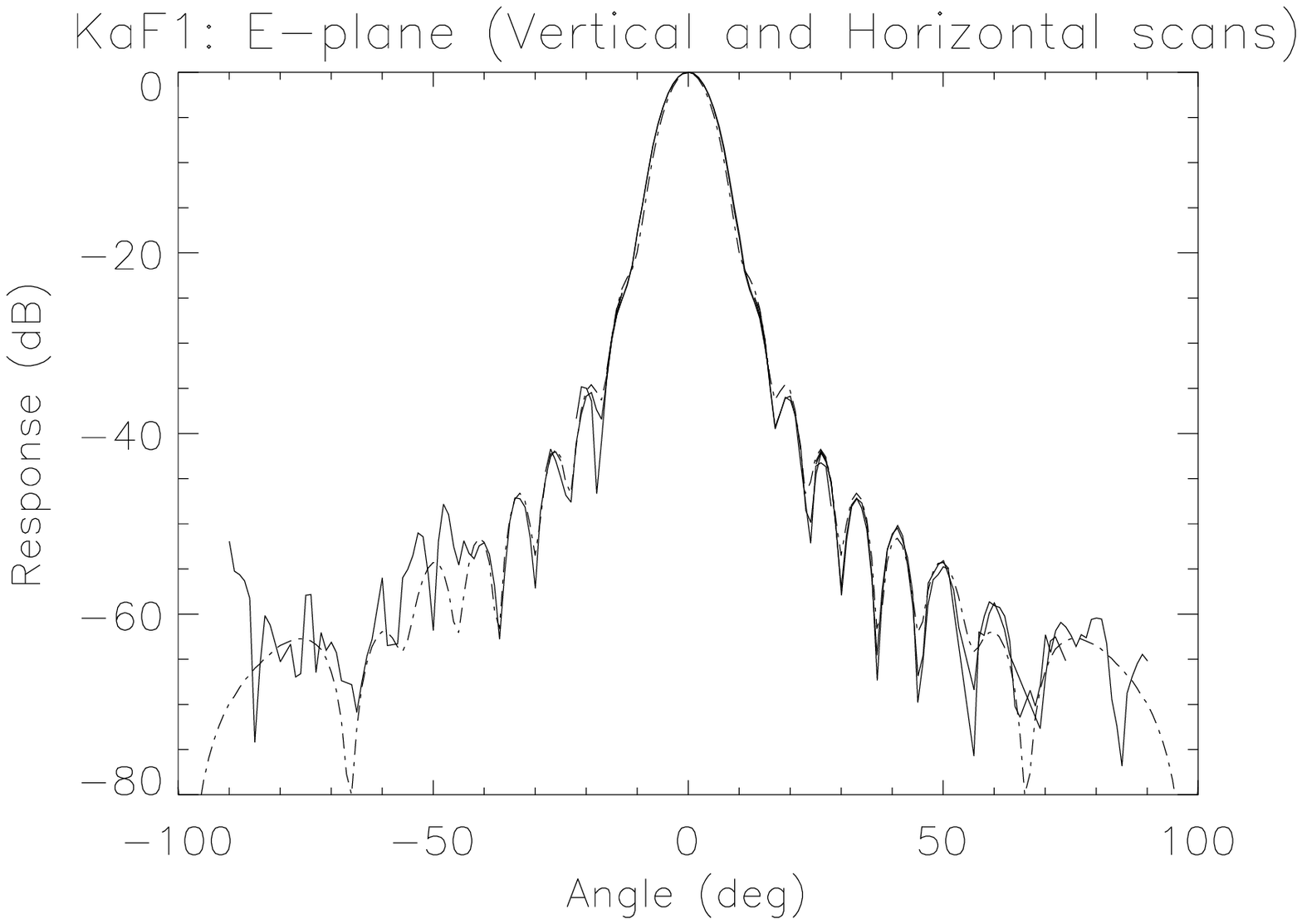} 
\plotone{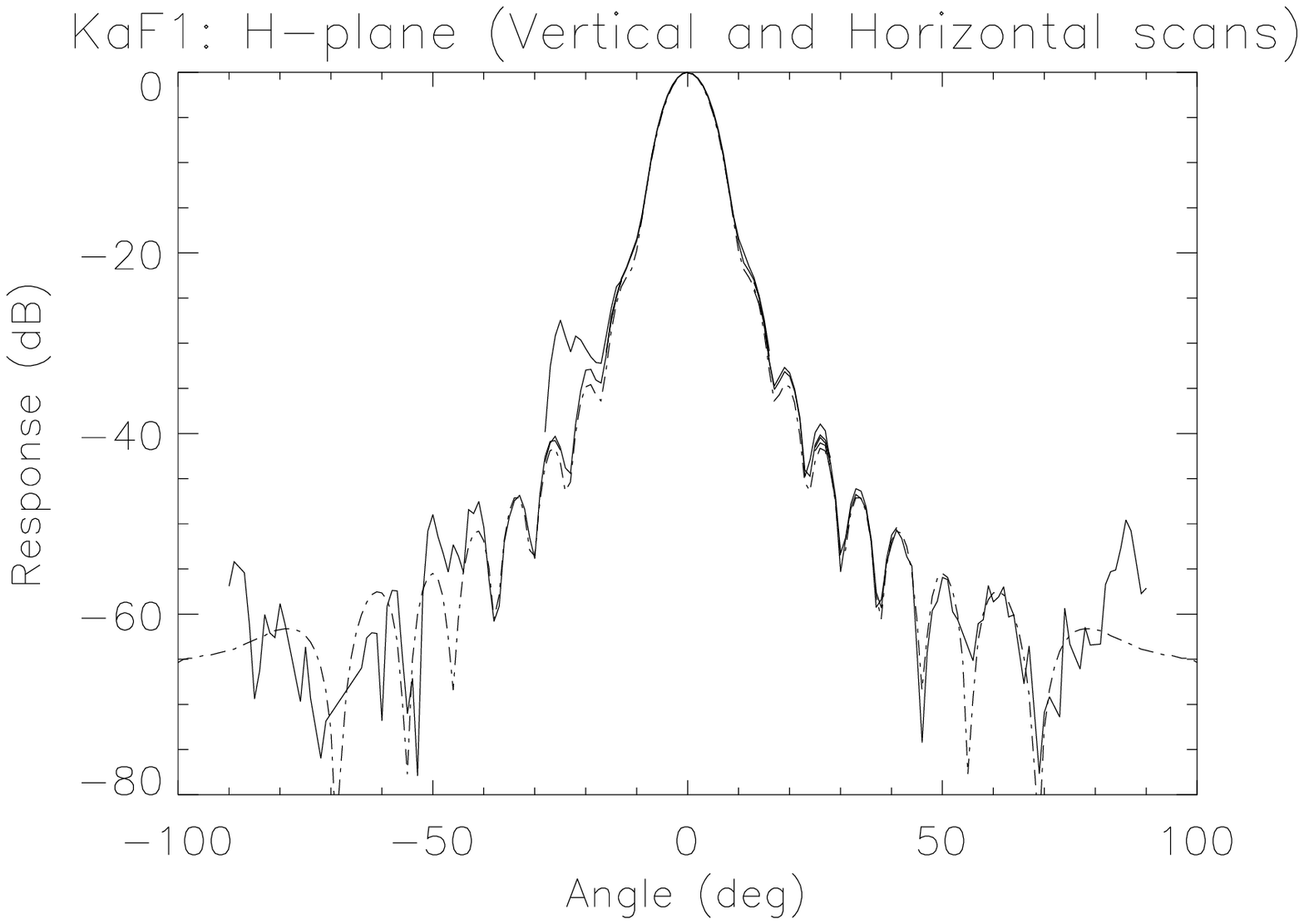} \\
\plotone{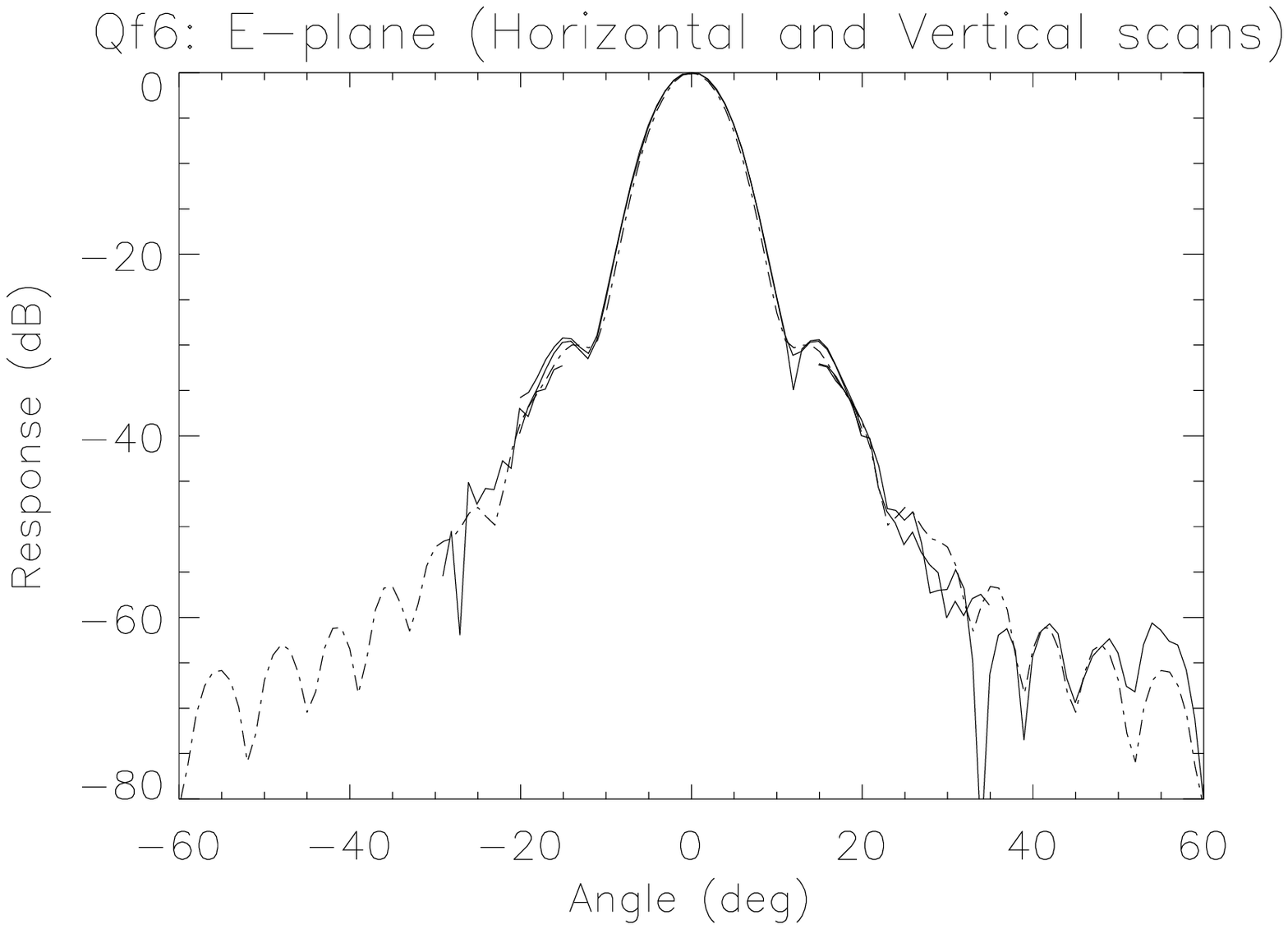} 
\plotone{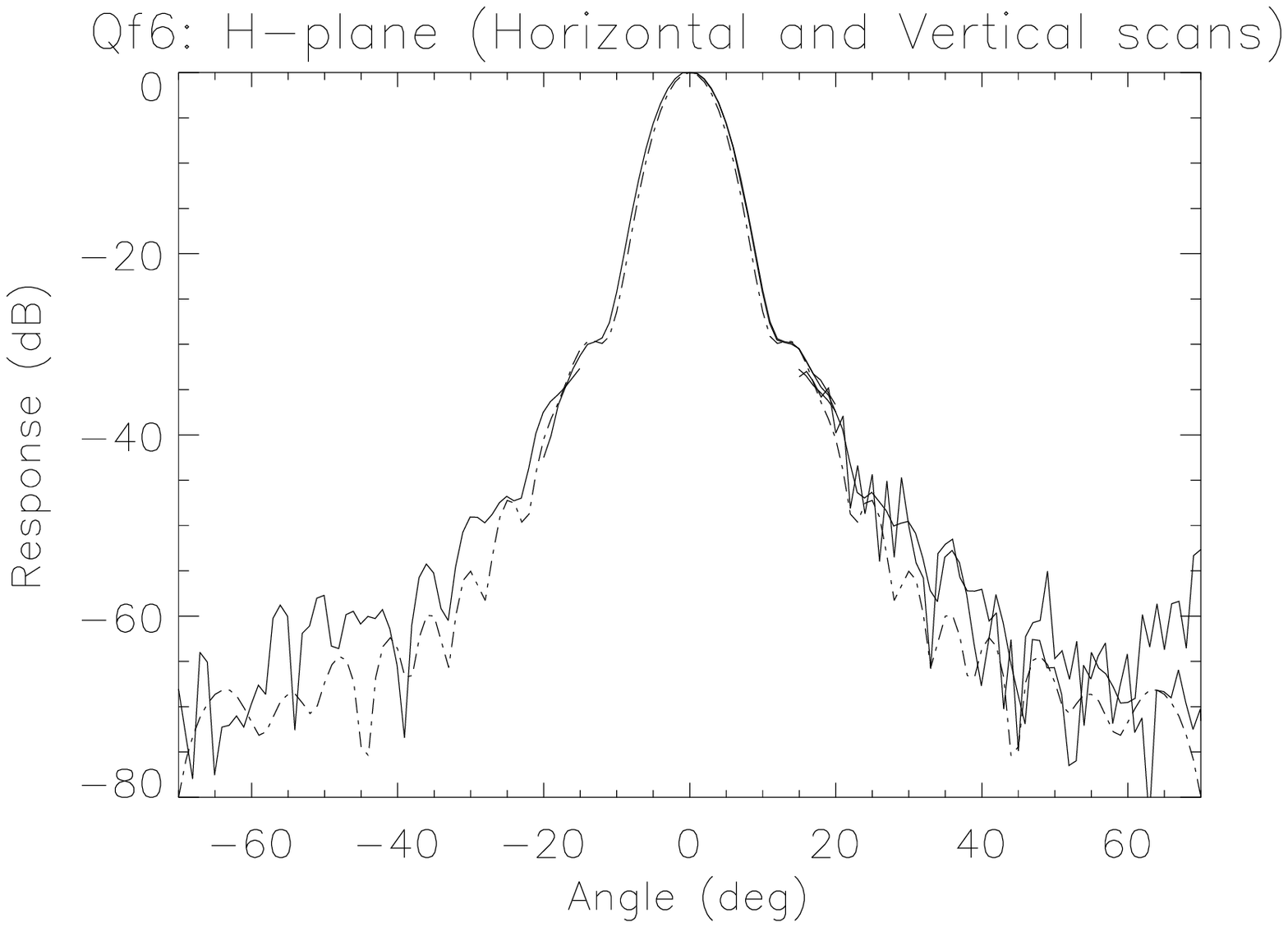} \\

\plotone{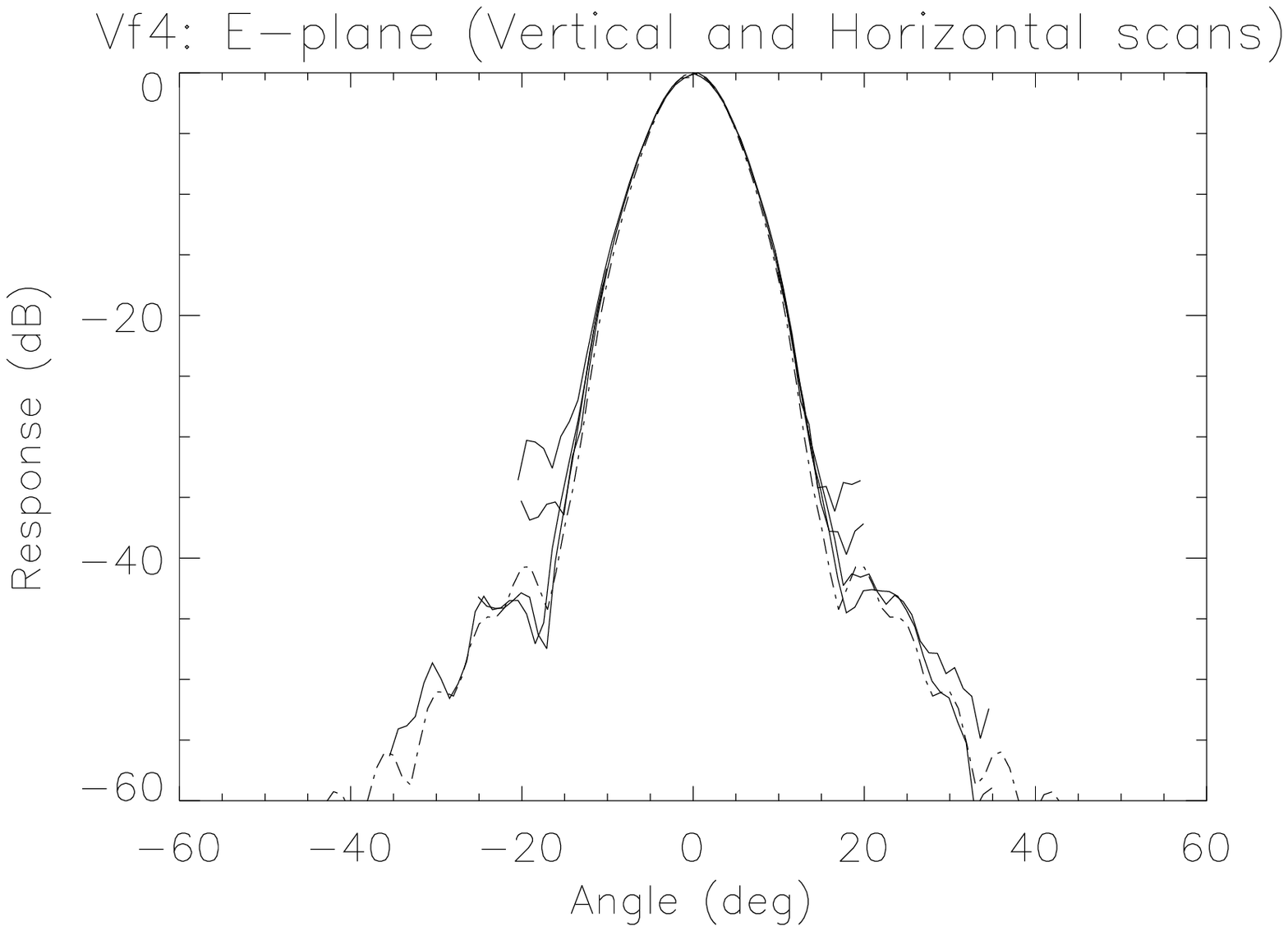} 
\plotone{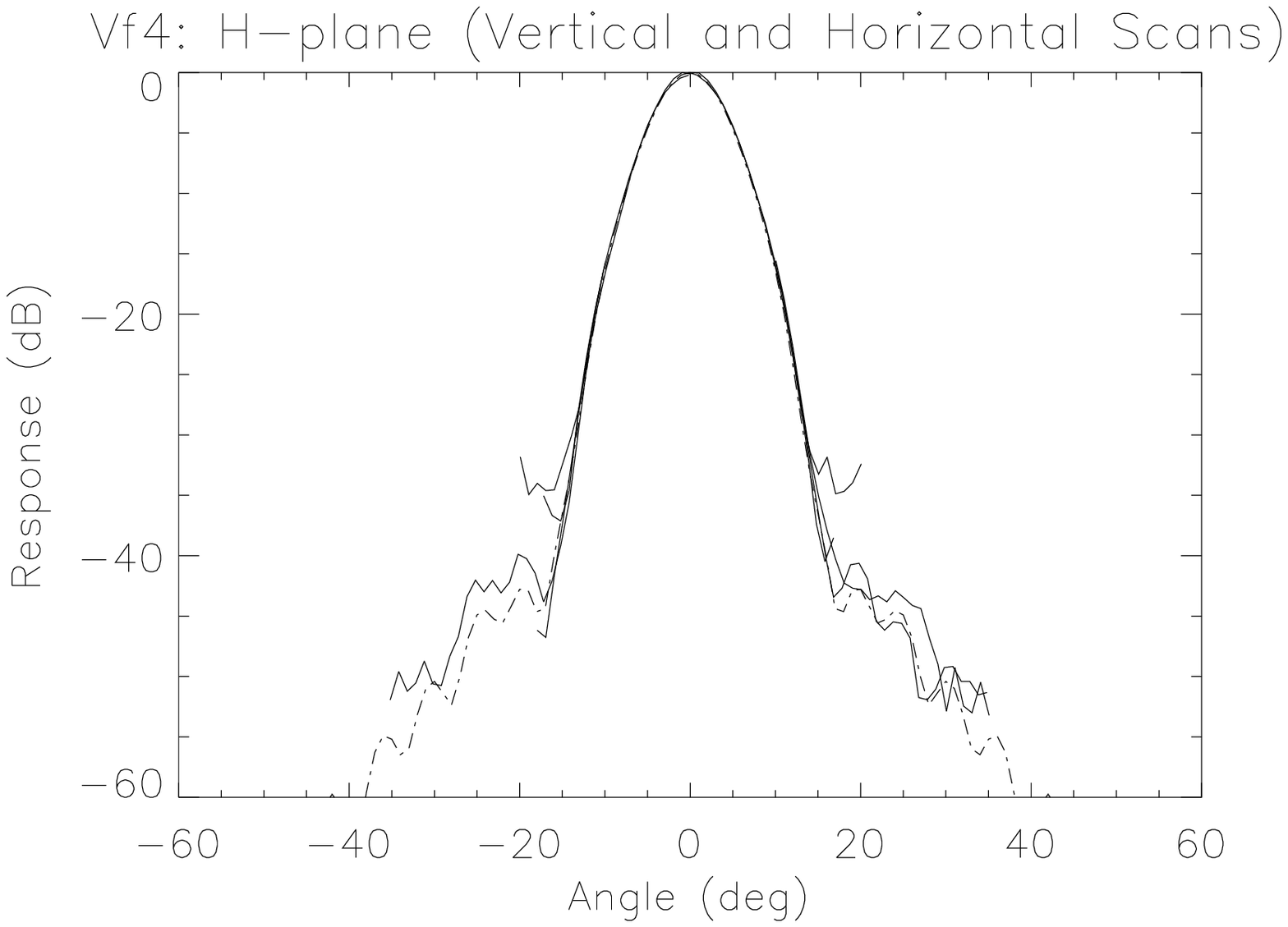} \\
\plotone{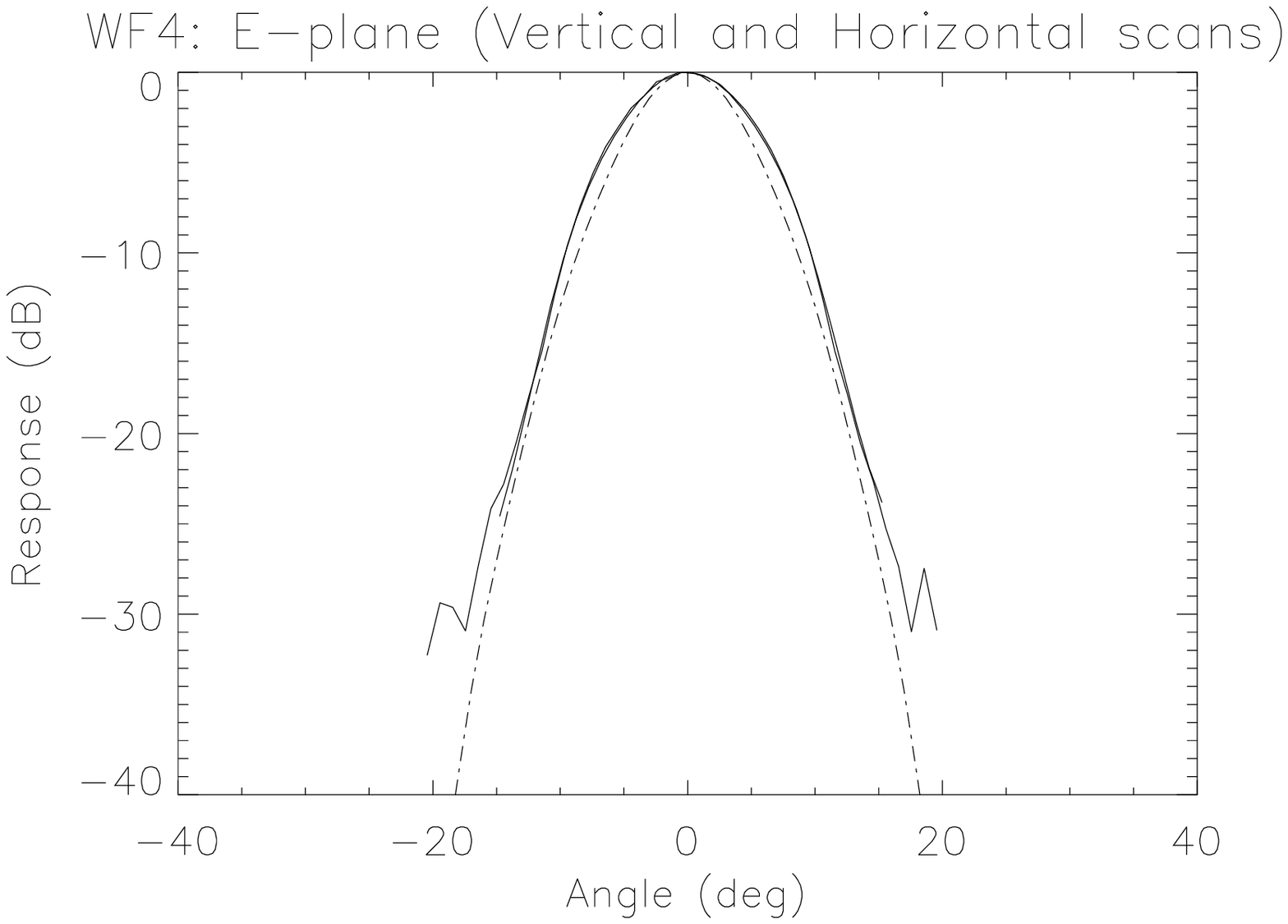}  
\plotone{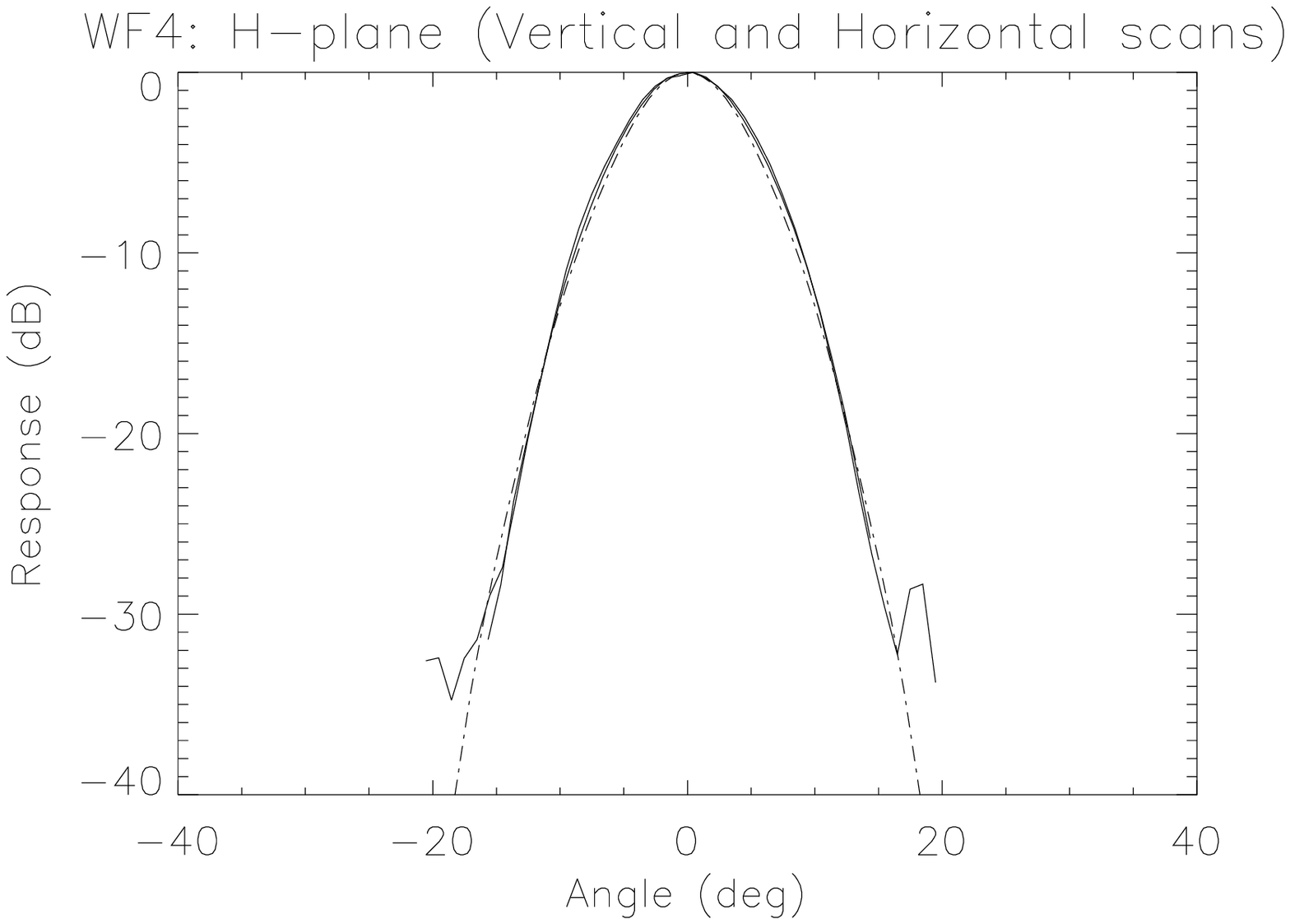} 
\caption{\small E and H-plane beam patterns for five feed horns.  
In these plots,
the solid lines are measured responses and the dashed curves are predicted
beam patterns, calculated as in section \protect{\ref{section:predictions}}.
With the exception of W-band, these beams were mapped in two stages: 
a first, inner map of the beam peak without an RF amplifier, and a
second, outer map of the side-lobes 
with the amplifier installed. These measurements are matched at their 
overlaps.  Each plot here overlays two scans: one vertical with the feed
scanning from the sky (positive angles) down into the roof (negative angles),
and one horizontal with the feed scanning from the source to the
horizon at a cross elevation of about $80\deg$.  Discrepancies in
the far side-lobes of these scans are evidence of reflections in the
range. The little ``legs'' at -30 dB result from the noise floor of the
main beam scan.}
\labfig{patterns}
\end{figure*}

One pair of antenna patterns from each band is shown in Figure 
\fig{patterns}, together with the beam pattern predictions for that horn 
design at that frequency.  The peak response of each beam has been
normalized; there is not an absolute calibration of the feed gain.  

The agreement between the theoretical and measured patterns is
excellent, particularly
in the central lobe.  A disagreement indicates an
imperfection in the horn. In Wf4, for example, the E-plane plot shows
a slight broadening near the beam peak, which indicates a second small
amplitude mode propagating through the feed.  (This
discrepancy was judged not large enough to disqualify the horn from
flight.)       

\begin{figure*}[ht]
\epsscale{1.0}
\plotone{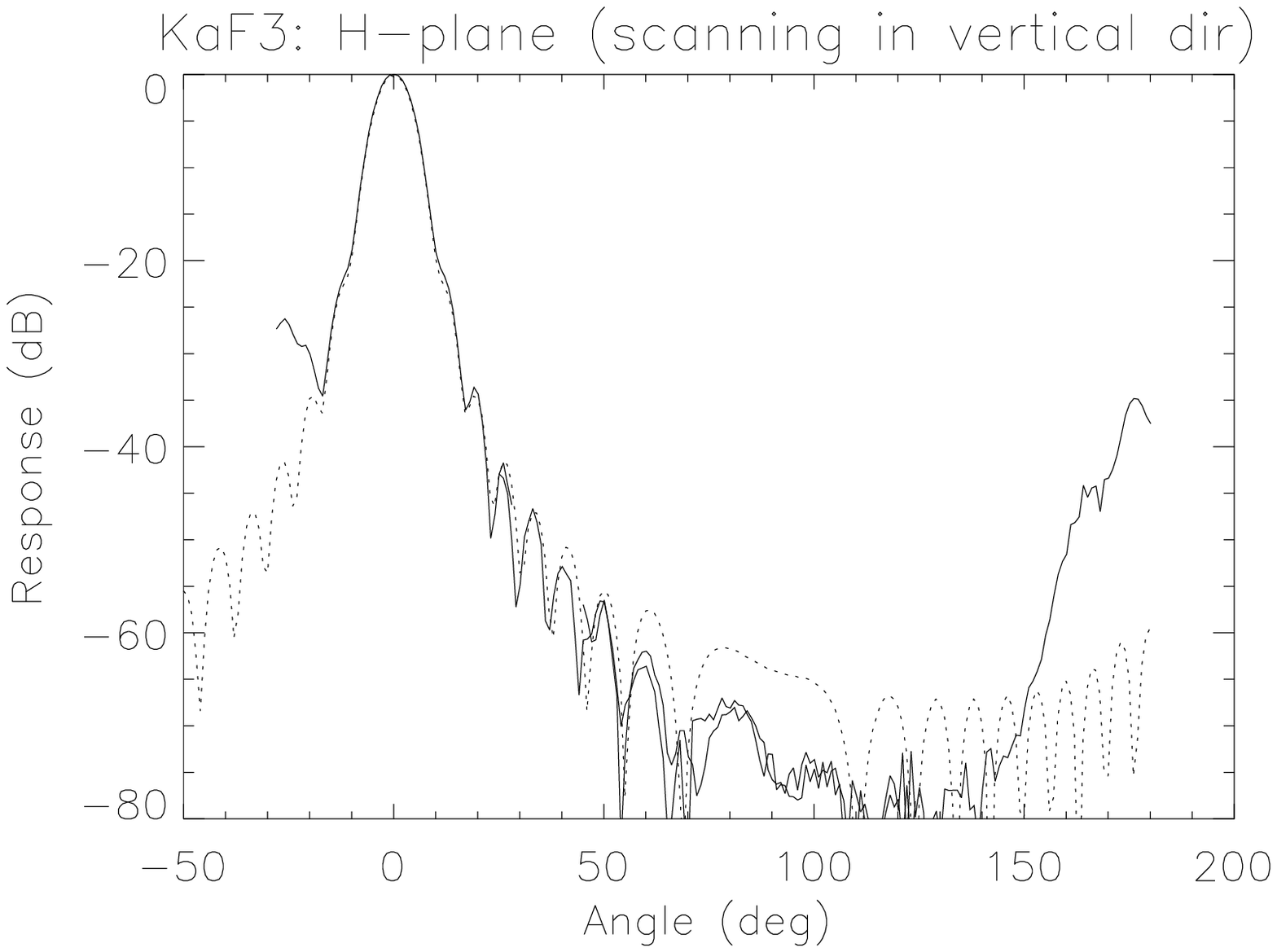}
\caption{\small Two deep sidelobe measurements of a Ka band horn.  Beyond 150
degrees, the horn looks below the horizon, and sees reflections from the
ground. Below -25 degrees, another unwanted reflection dominates.}
\labfig{ka_deep}
\end{figure*}

The feed's receiving pattern when it looks far away from the
source, at angles $>90\deg$ is also measured.  
The predicted
antenna pattern is extremely low in this region ($-60$ to $-90$ dB below
beam peak), so the measurement is more difficult, and sensitive to
reflections.  
Because the CMB anisotropy signal is so faint compared with the
brightest objects in the sky, we are compelled to understand this 
ordinarily ignorable section of the antenna pattern.
One such measurement of a Ka-band horn is shown in
Figure \fig{ka_deep}.  Beyond about $90\deg$, there is no longer
close agreement between the predicted and measured beam patterns.
This makes sense, when the horn is turned further than $90\deg$
away from the source there is an additional conducting
boundary, the {\it outside\/} of the horn, which becomes
important, and which is not included in the predictions.  In terms of
geometric optics, once the horn rotates beyond $90\deg$ from the
source, its aperture is in the shadow of the horn itself.  (It is tempting
to explain the low signal around $90\deg$ in Figure \fig{ka_deep}
purely in terms of this shadow, but that is an oversimplification. In
contrast, the E-plane scan is {\it higher} than predictions in the
same region.)  The feed's outer
surface is not included in the beam predictions, though, because 
the response was measured to be small, we know modifications 
to the model are unnecessary for our purposes. 

The predicted strength of the cross-polar antenna response varies
strongly across the band for each feed.  In each band, the cross-polar
antenna pattern is four-lobed, with the four lobes together
illuminating an angular region comparable to the co-polar beam
area. The predicted ratios of maximum cross-polar antenna gains to
maximum copolar gains are: in K-band, -31 dB, in Ka-band -39 dB, in
Q-band -39 dB, in V-band -33 dB, and in W-band -44 dB. In other words
in K-band the strongest cross-polar pickup for a source in the
band is 31 dB weaker than the peak copolar response at the same
frequency.  Cross-polar pickup depends strongly on frequency within
each band; the reported cross-polar sensitivity is the maximum
value for each feed.  Manufacturing defects will tend to increase
any cross-polar contamination.  

For {\sl MAP}, the feed's cross-polar signal is
dominated by mode-mixing in the OMT in all but K band \cite{Jarosik03}. 
The OMTs misdirect polarized radiation at $<-25~$ dB level, 
while the cross-polarization in
the feeds is predicted to be below $<-30~$ dB.  As
the OMTs limit the polarization separation, there was
no reason to probe the cross-polarized response of the horns deeper
than OMT level.  In K and Ka bands, the configuration of the optics has a
greater affect on the cross-polar response than either the feed or the OMT.
In terms of CMB measurements, even the polarization
mixtures induced by the OMTs will be an ignorably small contribution
to the error in {\sl MAP's} polarization measurement.

After the individual feed measurements, the feeds were installed in {\sl MAP}'s
focal planes and the entire optical system (with reflectors) 
was mapped in an 5706 electromagnetic anechoic chamber facility 
at NASA/GSFC called the GEMAC. The GEMAC is based on MI Technologies positioners
and 5706 antenna system. The receivers and microwave sources are made by
Anritsu. The GEMAC can map all {\sl MAP's} bands with 1~kHz resolution plane
wave source to  -50~dB from the peak response over almost 2$\pi$~sr with
an absolute accuracy of $<0.5$~dB. 
The resulting agreement between the predicted and main beam response 
(Page 2003 {\it et al.}), including absolute gain, is further evidence 
that the feeds met specifications. In all cases, the Princeton and GEMAC
beam profiles were consistent.
 
\section{Measurements of the reflections}
\label{section:reflections}

The {\sl MAP} feeds have $\le -25$ dB reflections across their bands, 
as measured with a Hewlett Packard 8510C 1-110~GHz network analyzer. It is 
necessary to connect the 
horn to the network analyzer
via a straight rectangular-to-circular waveguide transition (rather than
an OMT with one arm terminated), since the OMTs typically have much
higher reflection coefficients than do the feeds. The OMTs are
characterized separately.  As {\sl MAP} is a differential instrument,
an imbalance in its inputs leads
to an offset in which one input always appears hotter. This offset is 
present even if there is no celestial signal. A
large offset, in turn, leads to a higher $1/f$ noise knee in the
radiometer \cite{Jarosik03}. Reflections from the
OMT/feed assemblies can produce offsets through two effects.
In the first, noise emitted from the HEMT
amplifiers' inputs is reflected by the OMT/feed back into the 
radiometer with slightly different magnitudes. The offset from this effect is
roughly $ T_{offset} \sim (T_{amp} - T_{ant})(\alpha_A - \alpha_B)$
where $T_{amp}$ is the effective temperature of the noise power emitted from the
input of the radiometers HEMT amplifiers, and $\alpha_A$ and $\alpha_B$ 
are the power reflections coefficients for the OMT/feed assemblies 
connected to the two inputs, and $T_{ant}\approx 3~$K. 
For W-band, $T_{amp} \approx 100 K $.  
If $\alpha_A$ and $\alpha_B$ are both $\approx -25~$dB and are matched
to about 10\%, $T_{offset} \approx 50~$mK. 
In the second effect, the reflections produce coherent crosstalk between
the two HEMT amplifiers. Ideally, each of the two input HEMT amplifiers
sees the sum of two incoherent radiation fields, one from each side of the 
satellite. When the input noise from one HEMT reflects off the OMTs/feeds,
it reenters both arms of the differential receiver coherently and produces
an offset of $ T_{offset} \sim ({T_{amp} - T_{ant})\sqrt{\alpha} \beta \gamma}
\approx 0.2~{\rm K}$
where $\beta \approx 0.2$ is a correlation coefficient that indicates
the degree to which the noise emitted from the input of the HEMT is correlated
to the amplified noise at the output of the amplifier, 
$\alpha\approx\alpha_A\approx\alpha_B$, 
and $\gamma \approx 0.15$ (W-band) is a factor to allow
for the bandwidth averaging of the effect 
from the different path lengths the signals take before being recombined.
 
Samples of the measured reflection strengths are shown in Figure 
\fig{reflections} together with predictions. The network analyzer measures
a single mode and was calibrated for a rectangular
waveguide. The feed reflections are an extremely sensitive
indicator of manufacturing defects. 

\begin{figure*}[ht]
\epsscale{0.65}
\plotone{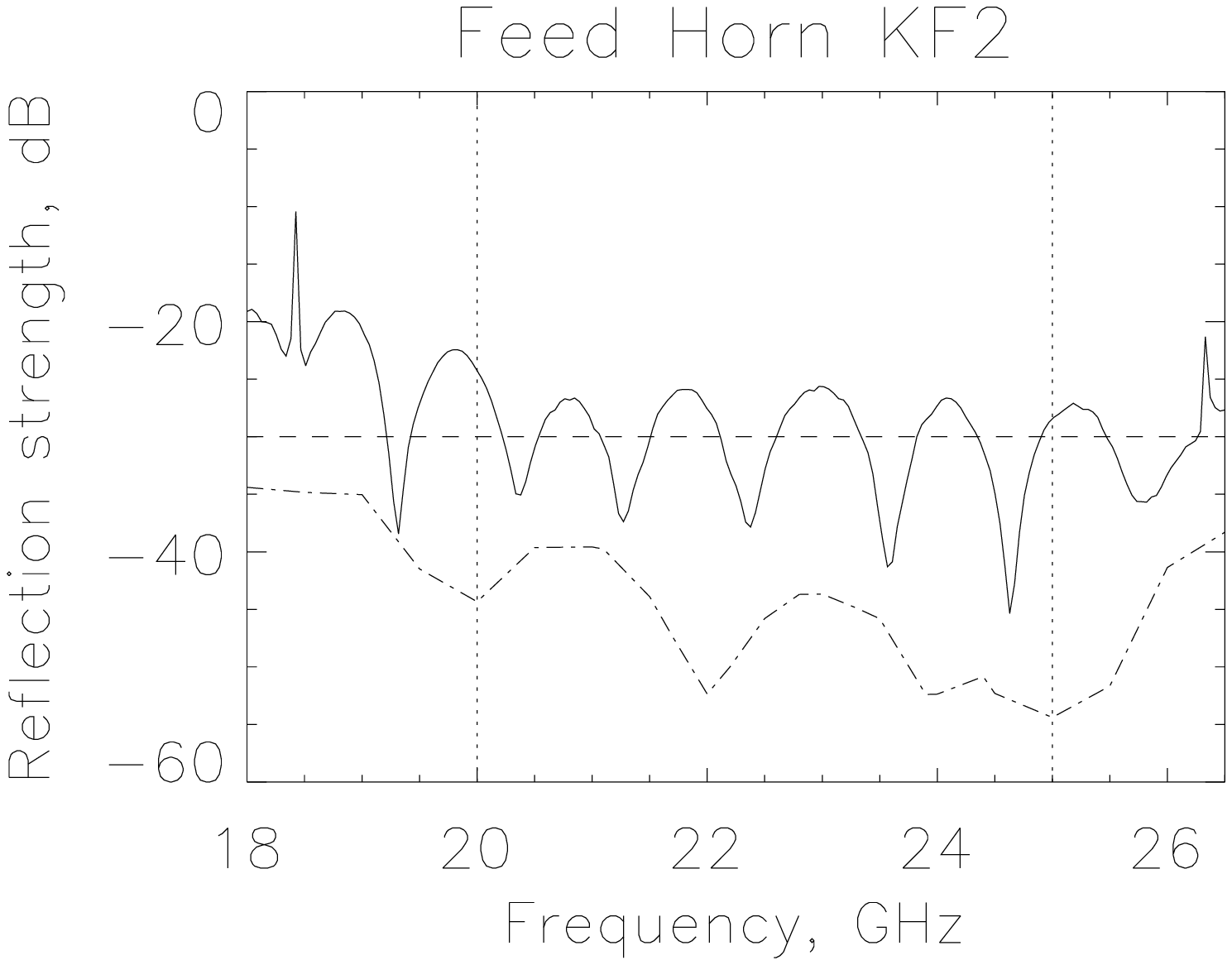}  
\plotone{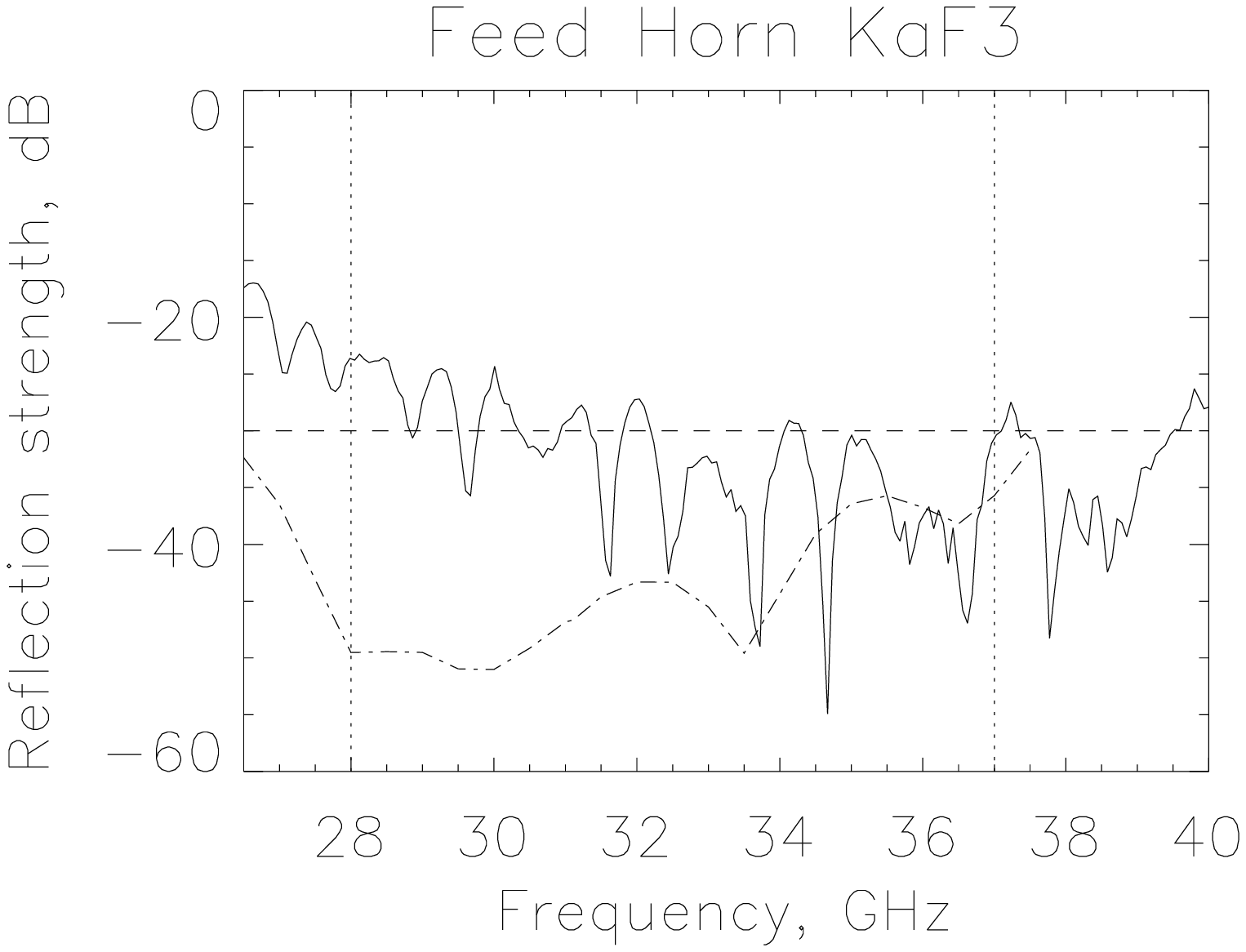} 
\plotone{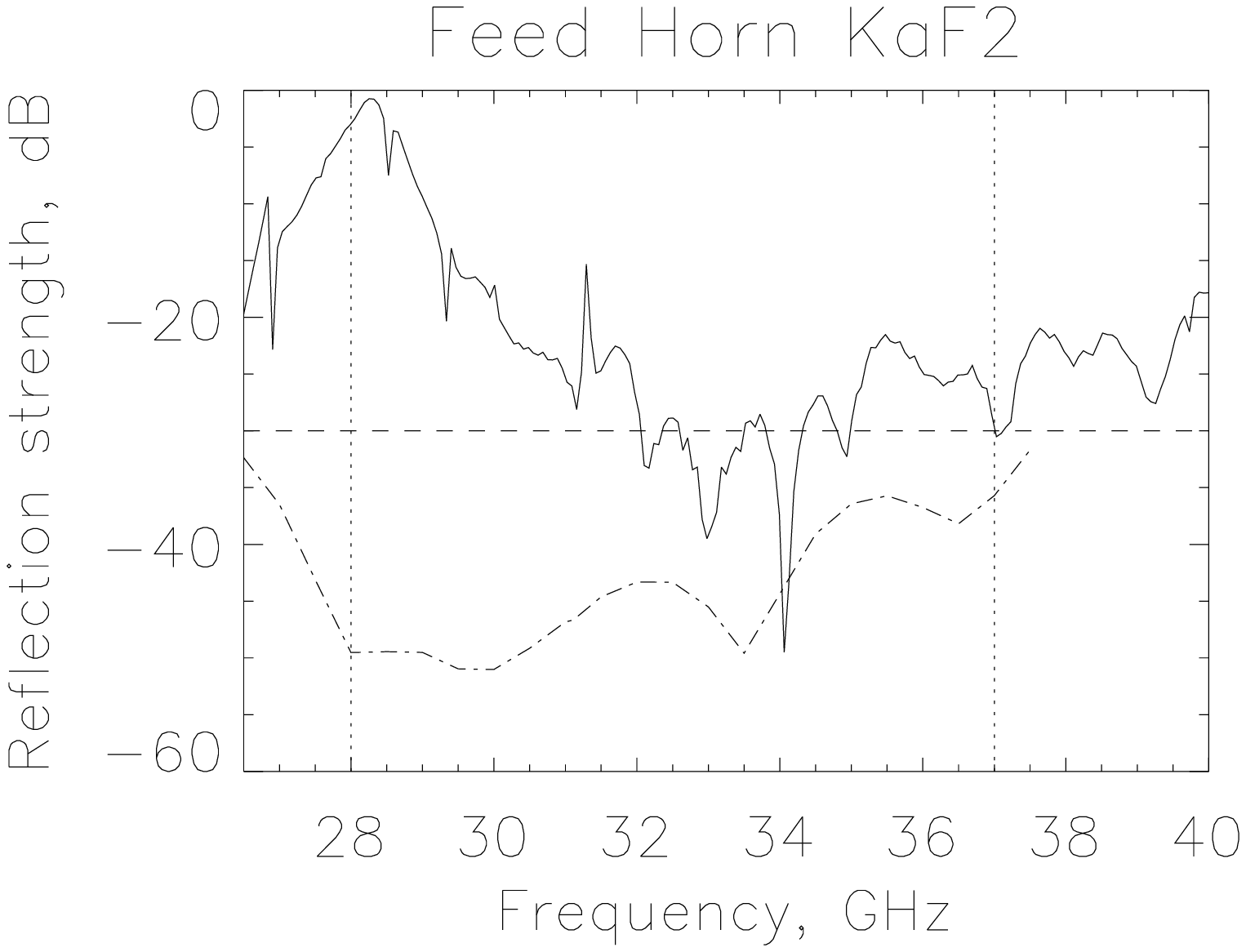} \\
\plotone{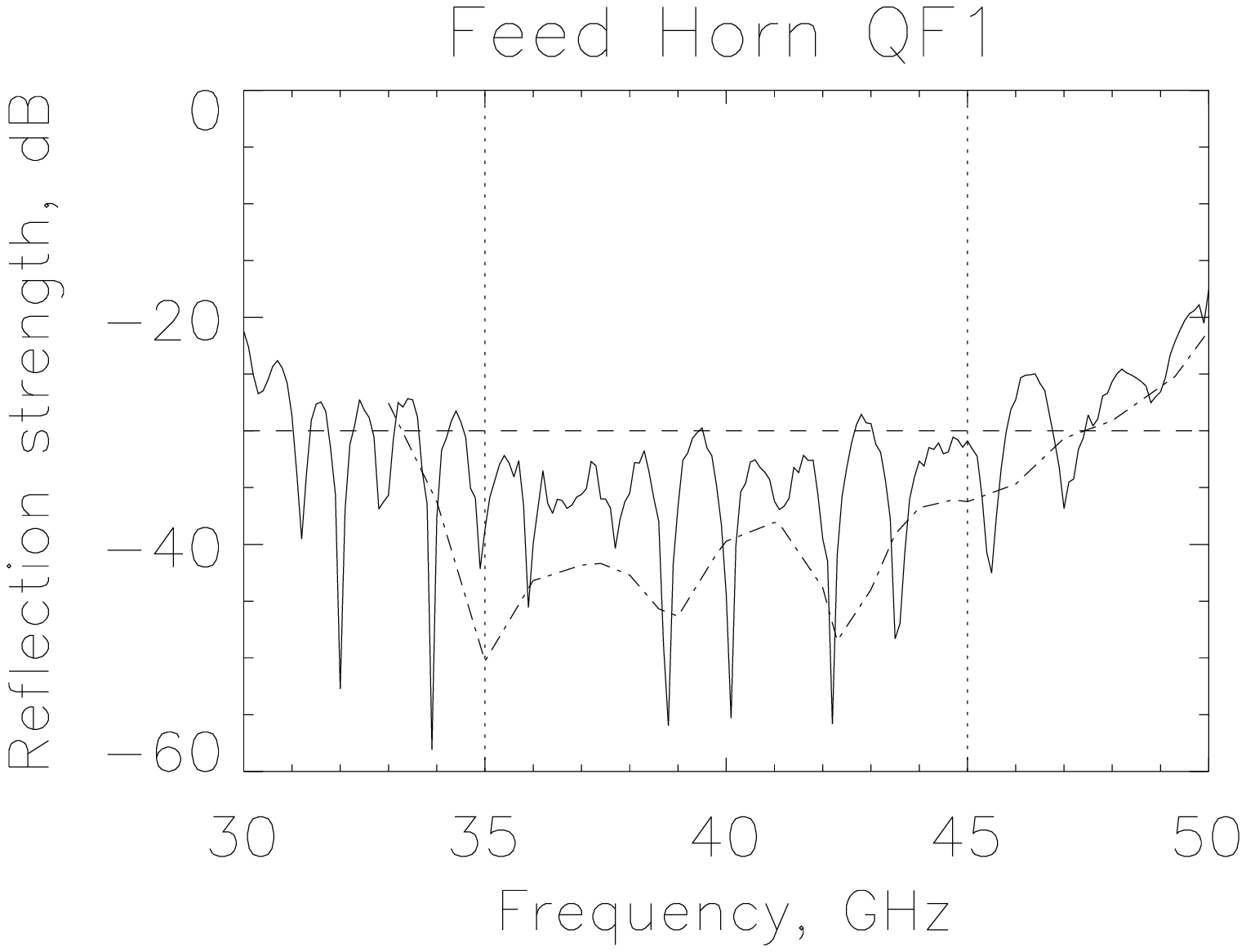} 
\plotone{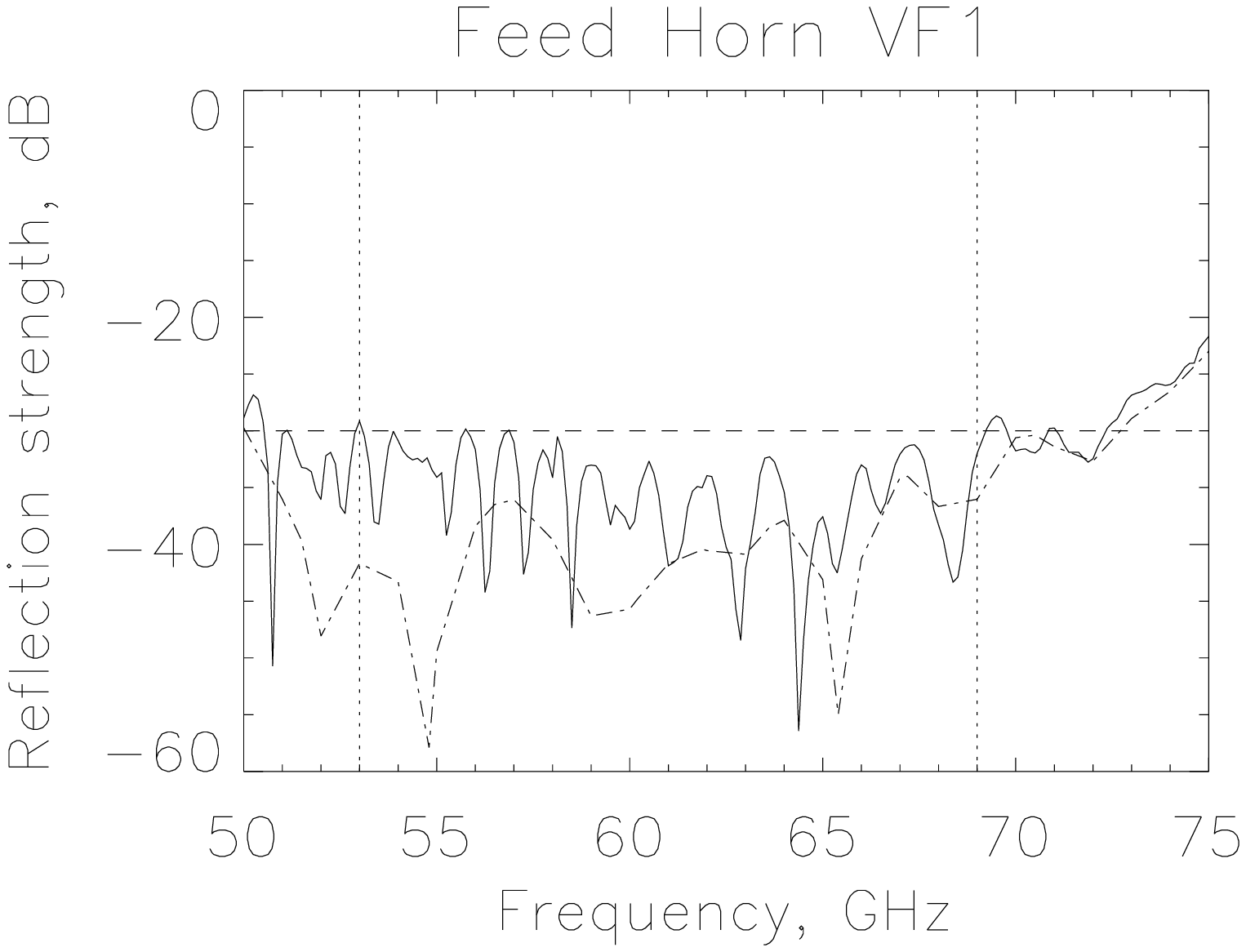} 
\plotone{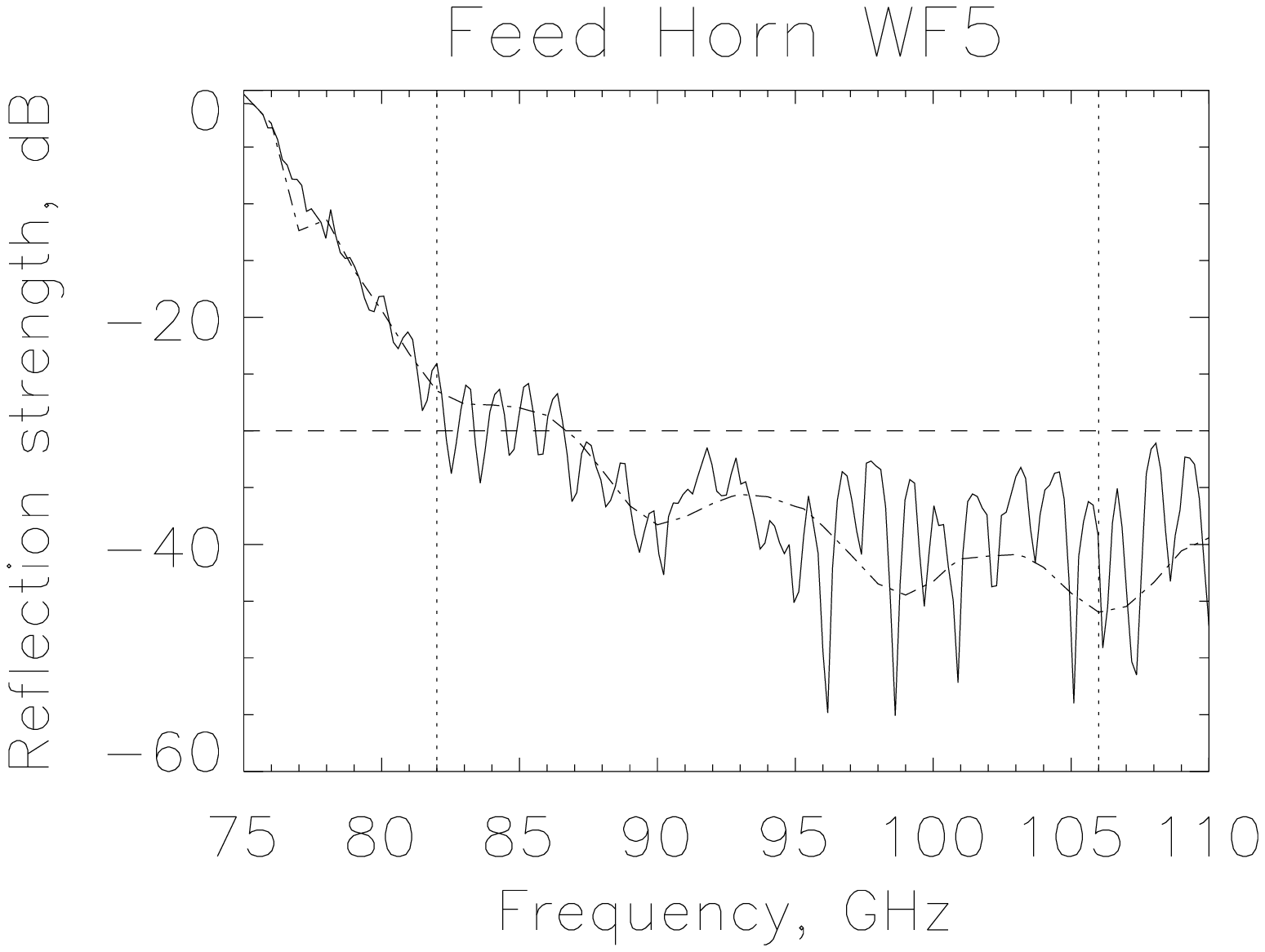} 
\caption{\small Predicted and measured reflected power for a selection
of the {\sl MAP} feed horns. The
measured and predicted reflections are solid and dashed respectively.
The reflected power is $10\log_{10}|S_{11}|^2$ where $S$ is the
scattering matrix.  The vertical dotted
lines mark {\sl MAP radiometer} band edges, and the horizontal 
dashed line provides a
reference at -30 dB.  The measured reflection strengths show the
combined effect of the rectangular-to-circular waveguide transition
and the feed.  (Predictions include only the feed.)  In particular,
the small regular oscillations in all the measured reflections
correspond to the rectangular-to-circular transition length. 
In K band, the return loss is dominated by the transition; in W band it
is dominated by feed. Feed
KaF2 has a shallow spiral scratch inside its throat, an accident of
fabrication. (KaF2 was not installed on the satellite.) }
\labfig{reflections}
\end{figure*} 

\section{Measurements of the Loss}

The loss in an ideal corrugated feed \cite{clarr_75} is small 
and often negligible;
consequently there are not many measurements of it. If
the loss is not balanced between the two sides of the differential 
radiometer, the output will have an offset. If the temperature of the
feeds varies significantly, the offset will vary and possibly
introduce a signal that confounds the celestial signal. 
For loss in the feeds, the offset between the $A$ and $B$ sides is
$T_{offset}\approx\epsilon_AT_A -\epsilon_BT_B$ where $\epsilon_A$ and
 $\epsilon_B$ are the feed emissivities. For example,
if $T_{offset}=1~$K with the feeds at $T_A=T_B=100$~K, the
temperature of the feeds must be stable to 0.1~mK to avoid spurious
signals $>1~\mu$K. The spacecraft
instrumentation can detect instability at this level.

The emissivity of a room temperature feed was measured \cite{stu_thesis} to be
$\epsilon=0.11$ at 90 GHz, well above the modeled 
emissivity of $\epsilon=0.017$. 
The measurement was made by uniformly heating a feed that was
thermally isolated from a stabilized 90 GHz receiver. 
The emission temperature of the feed was compared to the emission
temperature of a stabilized load at a variety of feed temperatures.

During the cold testing (100~K) of the {\sl MAP} receivers plus feed
horns, it was found that in one case an offset of $T_{offset}=2.5~$K 
could be clearly attributed to imbalanced
emission from the feeds corresponding to $\epsilon_a-\epsilon_B=0.025$.
If the loss 
were metallic, one would expect the emissivity of one feed to be
$\epsilon=0.057$ at 100~K and the difference between two feeds to be significantly
less than this. The offending feed was replaced and the amplitude of the
offset was reduced. Based on the differential measurements, the loss in 
other feeds was considered acceptably low; however, we cannot be certain that the
theoretical performance was achieved for any feeds because the measurement
is intrinsically differential. The in-flight performance will be addressed in
a future paper. 

The large emissivities of the test feed and the ``flight feed'' that was
replaced were traced to improper gold plating that became clearly 
evident after one of the electroformed feed sections was sliced open. 
To guard against deterioration of the other electroformed feed tails,
they were continuously purged with nitrogen until launch.

\section{Conclusion}

The nature of CMB anisotropy, microkelvin fluctuations in a sky
containing sources many orders of magnitude brighter, requires 
close attention to properties of optical elements of any CMB
telescope. Ordinarily ignorable features, particularly far off-axis 
antenna pickup, can significantly effect measured temperatures.
For each optical component of {\sl MAP}, it is crucial to understand 
the full antenna pattern, emissivity, and any backscatter native to
that element.  

The {\sl MAP} feeds produce near-Gaussian beams with low reflections
across their bands.  The measured beam patterns match the predictions
strikingly well; in some cases measurement and theory agree to $-80~$dB.
No significant discrepancy between the predicted and measured beam
patterns was seen at angles $<90\deg$ for a well
manufactured feed; beyond 90 degrees, the calculation neglects an
important boundary term.  Similar accuracy was not needed for the
reflection predictions as reflections are dominated by the
OMT. However, measurement confirm that the horn reflection spectra are
low enough for {\sl MAP}'s purposes. The feed cross-polar pattern is
computable and negligible. For {\sl MAP} the cross-polar pattern is
dominated by the OMT and reflector configuration. The feed loss is low
in most cases, but a manufacturing problem led to a demonstrably
higher than expected emissivity in two out of fourteen electroformed 
W-band feed tails.

We have shown that corrugated feeds can be manufactured in a variety 
of shapes and that with detailed attention to the manufacture, the 
theoretical performance may be achieved. Future missions will
undoubtedly use corrugated structures because of their many benefits.

\section{Acknowledgments}

The design, building and testing of the {\sl MAP} feeds took over two
years. The Princeton machine shop was invaluable in this program,
especially Glenn Atkinson who spent a year on machining alone. Ken
Stewart and Steve Suefert kept NASA/GSFC's GEMAC indoor beam mapping
facility mapping for weeks at a time. Charles Sule inspected all of
the feeds and Alysia Marino assisted in the measurements. The modeling 
of the feeds was made possible by assistance
and code from YRS Associates: Yahya Rahmat-Samii, Bill Imbriale, and
Vic Galindo. Of particular note, YRS Associates derived the groove dimensions,
feed shapes, and computed the fed emissivity. 
This research was supported by the {\sl MAP} Project
under the NASA Office of Space Science. 
More information about {\sl MAP} may be found at http://map.gsfc.nasa.gov.

\end{document}